%% file: NMRB_main.tex
\documentclass[ASNA,twocolumn]{Definitions/nmrb/USG} 

\Newlabel{*}{*}

\usepackage{anyfontsize} %

\usepackage{xcolor}     
\graphicspath{{Definitions/nmrb/images/}}
\articletype{PREPRINT SUBMITTED TO NMR IN BIOMEDICINE}%

\received{xx xxxxx xxxx}
\revised{xx xxxxx xxxx}
\accepted{xx xxxxx xxxx}
\journal{NMR in Biomedicine}
\copyyear{2026}
\articledoi{10.1002/0000}  

\usepackage{mrsi_dataset_style}
\usepackage{nicefrac}

\usepackage{cuted}
\usepackage{pgf}          
\usepackage{caption} 
\usepackage{placeins} 
\usepackage{afterpage}
\usepackage{todonotes}

\begin{document}

\title{The 2024 MRSI Data Processing and Quantification Challenge Synthetic Dataset}
\author[1,2,*]{John T. LaMaster}[https://orcid.org/0000-0002-2149-771X]
\author[3,4,*]{Julian P. Merkofer}[https://orcid.org/0000-0003-2924-5055]
\author[5,*]{Dennis M. J. van de Sande}[https://orcid.org/0000-0001-6112-1437]
\author[6]{Brian J. Soher}[https://orcid.org/0000-0003-2750-7277]
\author[7]{Bernhard Strasser}[https://orcid.org/0000-0001-9542-3855]
\author[8]{Chao Ma}[https://orcid.org/0000-0002-2016-8084]

\authormark{LaMaster \textsc{et al.}}
\titlemark{2024 MRSI Challenge Dataset}

\address[1]{\orgdiv{Munich Institute of Biomedical Engineering, }\orgname{Technical University of Munich, }%
\orgaddress{\state{Bavaria, }\country{Germany}}}

\address[2]{\orgdiv{School of Computation, Information, and Technology, }\orgname{Technical University of Munich, }%
\orgaddress{\state{Bavaria, }\country{Germany}}}

\address[3]{\orgdiv{Department of Electrical Engineering, }\orgname{Eindhoven University of Technology, }%
\orgaddress{\state{North Brabant, }\country{The Netherlands}}}

\address[4]{\orgdiv{Center for Image Sciences, }\orgname{University Medical Center Utrecht, }%
\orgaddress{\state{Utrecht, }\country{The Netherlands}}}

\address[5]{\orgdiv{Department of Biomedical Engineering, }\orgname{Eindhoven University of Technology, }%
\orgaddress{\state{North Brabant, }\country{The Netherlands}}}

\address[6]{\orgdiv{Center for Advanced MR Development, Department of Radiology, }\orgname{Duke University Medical Center, }%
\orgaddress{\state{North Carolina, }\country{United States of America}}}

\address[7]{\orgdiv{High-field MR Center, Department of Biomedical Imaging and Image-guided Therapy, }\orgname{Medical University of Vienna, }%
\orgaddress{\state{Vienna, }\country{Austria}}}

\address[8]{\orgdiv{Yale Biomedical Imaging Institute, Department of Radiology and Biomedical Imaging, }\orgname{Yale School of Medicine, }%
\orgaddress{\state{Connecticut, }\country{United States of America}}}

\address[9]{\llap{\textcolor{white}{\textsuperscript{4}}}\llap{\textsuperscript{*}}These authors contributed equally.}

\corres{\email{john.lamaster@tum.de}, \email{j.p.merkofer@tue.nl}, \email{d.m.j.v.d.sande@tue.nl}, \email{brian.soher@duke.edu}, \email{bernhard.strasser@meduniwien.ac.at}, and \email{chao.ma.cm2943@yale.edu}.}

\fundingInfo{Nothing to declare.}


\abstract[ABSTRACT]{
Synthetic data are central to magnetic resonance spectroscopy (MRS) method development because they provide known ground truths for software validation, reproducible benchmarking, and machine- and deep-learning training. In magnetic resonance spectroscopic imaging (MRSI), useful synthetic data must capture spatially varying anatomy, field inhomogeneity, nuisance signals, and measurement-induced effects. The 2024 MRSI Data Processing and Quantification Challenge Synthetic Dataset is a simulated 3T brain free induction decay (FID)--MRSI resource with ground-truth metabolite maps that was developed as a controlled and reproducible testbed for MRSI processing and quantification methods. 

Subject-specific simulations were generated from anatomical images and field maps derived from Human Connectome Project subjects. Tissue masks, quantum-mechanically simulated metabolite basis functions, \textit{in vivo}-derived macromolecular components, Bloch-simulated post-WET residual water, non-rigidly registered \textit{in vivo} lipid signals, a spectral baseline, $B_0$-dependent frequency shifts, Voigt lineshape variations, and complex Gaussian noise were combined in a forward model. Water and lipid signals were synthesized on a high-resolution grid and Fourier-truncated to the final echo-planar spectroscopic imaging grid to model the finite spatial point spread function and its resulting spatial leakage. 

The resource comprises 24 training and 8 testing subject-level datasets containing contaminated FID--MRSI data, anatomical images, $B_0$ maps, metadata, and ground-truth metabolite and component signals. Forward-model parameters are fully documented, including tissue-specific metabolite concentrations and relaxation times, macromolecular amplitude ratios, the water model, noise, and dataset-specific field-inhomogeneity ranges. This controlled synthetic benchmark supports development and comparison of MRSI processing, nuisance-signal removal, and metabolite-quantification methods, while enabling reproducible evaluation of methods using known ground truth. More broadly, the resource illustrates how fully characterized synthetic data can support rigorous evaluation of computational methods when experimental ground truth is difficult or impossible to obtain.}

\abbr{
CRLB, Cram\'{e}r--Rao lower bound;
CSF, Cerebrospinal fluid;
DL, Deep learning;
EPSI, Echo-planar spectroscopic imaging;
FID, Free induction decay;
FWHM, Full width at half maximum;
GM, Gray matter;
HCP, Human Connectome Project;
HLSVD, Hankel Lanczos singular value decomposition;
ML, Machine learning;
MM, Macromolecules;
MRS, Magnetic resonance spectroscopy;
MRSI, Magnetic resonance spectroscopic imaging;
MRSinMRS, Minimum reporting standards for \textit{in vivo} MRS;
MRSsynMRS, Synthetic MRS Data Reporting Standards;
NIfTI, Neuroimaging Informatics Technology Initiative;
NIfTI--MRS, NIfTI format for magnetic resonance spectroscopy;
PSF, Point spread function;
SLR, Shinnar--Le Roux;
SNR, Signal-to-noise ratio;
SPM, Statistical parametric mapping;
SyN, Symmetric normalization;
TE, Echo time;
TR, Repetition time;
WET, Water suppression enhanced through $T_1$ effects;
WM, White matter.}

\contributed{John LaMaster, Julian Merkofer, and Dennis van de Sande contributed equally to this study.}


\maketitle

\acresetall

\clearpage
\section{Introduction}

\Ac{mrsi} extends \ac{mrs} from localized spectra to spatial maps of neurochemical signal distributions. This spatial information is highly valuable for studying heterogeneous brain metabolism, but it also makes processing and quantification substantially more difficult than in single-voxel \ac{mrs}. In addition to spectral overlap, line broadening, macromolecular background, residual water, lipid contamination, baseline variation, and noise, \ac{mrsi} requires explicit consideration of spatial encoding, tissue fractions, k-space sampling, the spatial response function, and spatially varying \(B_0\) and transmit/receive ($B_1^+$/$B_1^-$) fields \cite{LaMaster2026SyntheticReview,Kreis2021Terminology,Juchem2021B0Shimming}. As a result of these growing complexities, algorithm performance can vary across slices, tissue classes, metabolites, and artifact conditions.

A persistent limitation for \ac{mrsi} method development is the lack of realistic datasets with known metabolite ground truth. \textit{In vivo} data contain the biological and instrumental variability that algorithms must ultimately handle, but true metabolite maps are not directly observable. Synthetic data provide a complementary solution by allowing the relevant parameters and component signals to be specified explicitly. Recent community guidance emphasizes that synthetic \ac{mrs} data can support acquisition optimization, software validation, reproducibility, and ML/DL development, particularly when data are generated with transparent assumptions, realistic variability, standardized formats, accessible metadata, and complete reporting of simulation parameters \cite{LaMaster2026SyntheticReview,Clarke2022NIFTIMRS}. These principles are especially relevant for \ac{mrsi}, where benchmarking requires both spectral and spatial realism.

Existing simulation tools and digital phantoms have made important progress toward reusable synthetic \ac{mrs} resources. Basis-set and spectrum simulators such as Vespa, MRS-Sim, and synMARSS provide flexible ways to generate metabolite responses and \textit{in vivo}-like spectra under controlled parameter settings \cite{Soher2023Vespa,LaMaster2025MRSSim,VandeSande2026DigitalPhantom}. Reviews of ML in proton \ac{mrs} have also highlighted the need for large, well-described training and validation data when developing data-driven methods for denoising, artifact removal, reconstruction, and quantification \cite{VandeSande2023MLReview}. For \ac{mrsi}, however, a useful benchmark dataset must go beyond independent voxel-wise spectra: it should include anatomically structured metabolite maps, realistic nuisance signal propagation, \(B_0\)-dependent frequency and lineshape changes, and partial-volume effects arising from k-space truncation.

The dataset described here was created to address this gap. It was originally developed for the 2024 \ac{mrsi} Data Processing and Quantification Challenge\footnote{The challenge website is available at \url{https://bsoher.github.io/FittingChallenge2024/}.}, but the focus of this work is the dataset and its simulation strategy rather than the competition outcomes. It provides realistic \ac{fid}--\ac{mrsi} data, anatomical images, \(B_0\) maps, and ground-truth metabolite and component signals, organized around two sub-challenges that are described in Section~\ref{sec:benchmark}. The aim of this work is to describe the dataset in a form that makes it reproducible, not merely usable. Therefore, this work reports the numerical value of every parameter that enters the forward model, the exact composition of each released array, and the noise and field-inhomogeneity ranges of all 32 datasets, and the description is aligned with the proposed \ac{mrssynmrs} reporting framework.

\section{Materials and Methods}

\subsection{Dataset Design}
The dataset was designed as a synthetic 3\,T brain \ac{fid}--\ac{mrsi} resource with realistic spatial structure and known ground truth. The design followed four requirements. First, the simulated data should be challenging enough to expose practical failure modes in processing pipelines, especially in regions affected by strong \(B_0\) variation, residual water, lipid contamination, baseline variation, and low \ac{snr}. Second, the simulations should retain separable component signals and metabolite maps so that algorithms can be evaluated against known values. Third, data should be distributed in interoperable \ac{nifti} and \ac{nifti-mrs} formats, consistent with recent standardization efforts in the \ac{mrs} community \cite{Clarke2022NIFTIMRS}. Fourth, the simulation framework should be modular, allowing later versions to alter one source of variability without redesigning the full dataset.

The dataset consists of 24 training and 8 testing subject-level datasets. The testing partition is split into five datasets for sub-challenge 1 containing water and lipid nuisance signal contamination, denoted TestSub1--5, and three datasets for sub-challenge 2 without nuisance signals, denoted TestSub10--12. Both partitions were generated with the same source information and the same forward model. Because the water and lipid signals were deliberately scaled to mimic an \ac{fid}--\ac{mrsi} acquisition with very short \ac{tr} and \ac{te}, the nuisance signals in this dataset are stronger than in a typical \textit{in vivo} \ac{mrsi} experiment.

\begin{figure*}
\centering
\includegraphics[width=1.0\textwidth,keepaspectratio]{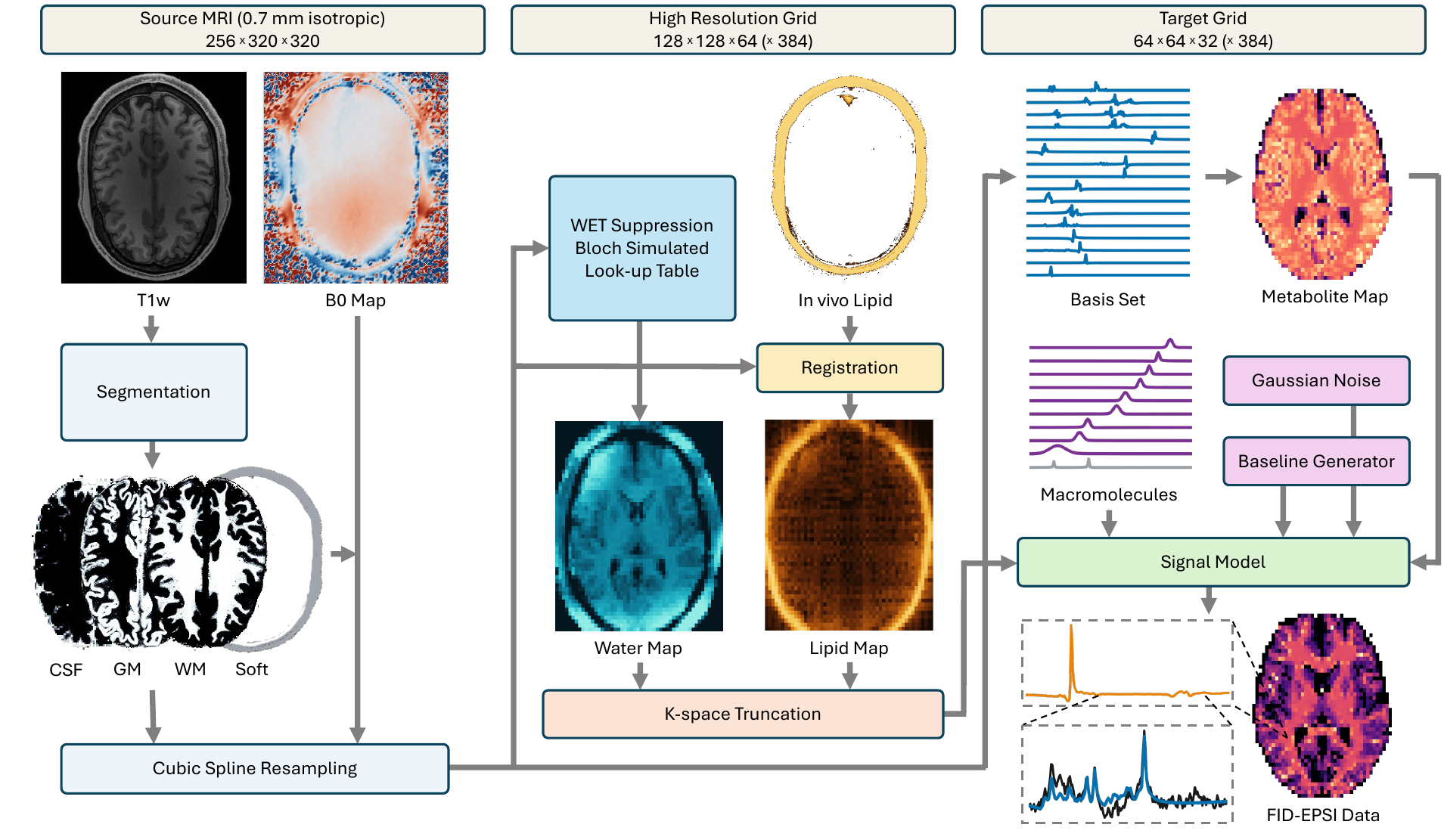}
\caption{Overview of the simulation workflow. Subject-specific anatomical images and \(B_0\) maps (left) provide the tissue-fraction maps $\Omega_\tau$, the field-dependent water-suppression efficiency, and the voxel-wise frequency offsets that drive the forward model. Water and lipid are synthesized on the high-resolution $128\times128\times64$ grid and Fourier-truncated to the target grid, in order to model the finite point-spread function to reproduce the spatial leakage. The metabolite, macromolecular, and baseline terms are built directly on the target $64\times64\times32$ grid and therefore do not exhibit Gibbs ringing. All components are summed with complex Gaussian noise to form the released \ac{fid}--\ac{mrsi} data, alongside the component-level ground truth. 
}
\label{fig:simulation_overview}
\end{figure*}

\subsection{Overall Signal Model}

For spatial location \(\mathbf{r}=(x,y,z)\) and sampling time \(t_n = (n+2)\,\delta t\), \(n = 0,\dots,383\), the synthetic \ac{mrsi} signal is:

\begin{equation}
    \begin{split}
    s(\mathbf{r},t) = {}& \underbrace{s_M(\mathbf{r},t)}_{\text{metabolites}} + \underbrace{s_B(\mathbf{r},t)}_{\text{\ac{mm}}}\\
    &+ \underbrace{\mathcal{T}\!\left\{s_W\right\}(\mathbf{r},t) + \mathcal{T}\!\left\{s_L\right\}(\mathbf{r},t)}_{\text{nuisance}}\\
    &+ \underbrace{s_{\mathrm{BL}}(\mathbf{r},t)}_{\text{baseline}} + \epsilon(\mathbf{r},t),
    \end{split}
    \label{eq:overall_signal_model}
\end{equation}
where \(s_M\), \(s_B\), \(s_W\), \(s_L\), and \(s_{\mathrm{BL}}\) denote the metabolite, macromolecular, residual-water, lipid, and baseline signals, \(\epsilon\) denotes complex Gaussian noise, and \(\mathcal{T}\{\cdot\}\) denotes the k-space truncation operator described in Section~\ref{sec:truncation}. This component-wise formulation follows current recommendations for synthetic \ac{mrs} data, which ask authors to document the basis set, amplitude model, lineshape model, nuisance components, noise model, and spatial assumptions \cite{LaMaster2026SyntheticReview}. It also enables component-level ground truth, which is essential for evaluating whether nuisance removal suppresses lipid and water signal without removing metabolite signal.

One aspect of Equation~\eqref{eq:overall_signal_model} deserves emphasis, because it differs from a na\"{i}ve reading of the workflow: \(\mathcal{T}\) acts only on the water and lipid terms. These are the components whose spatial leakage matters, and they alone are synthesized on the high-resolution grid. The metabolite, \ac{mm}, and baseline terms are synthesized directly on the target $64\times64\times32$ grid and are therefore free of Gibbs ringing.

All components share a common voxel-wise frequency modulation $e^{i2\pi\Delta f(\mathbf{r})t}$ induced by the $B_0$ map, and the metabolite and macromolecular terms additionally share non-Lorentzian lineshapes determined by the local $B_0$ inhomogeneity. Since $t$ begins at $\mathrm{TE}=2\delta t=1.66$\,ms rather than at zero, the frequency modulation also imparts a $\Delta f$-dependent first-order phase. No additional random phase is applied. Figure~\ref{fig:simulation_overview} summarizes the workflow, from the source anatomical images through the two synthesis grids to the released data volumes.

\subsection{Anatomical Templates, Tissue Masks, and Field Maps}
High-resolution $T_1$-weighted anatomical images and gradient-echo field maps from young adult \ac{hcp} subjects were used as templates \cite{VanEssen2013HCP}. The anatomical images are the \ac{hcp} 3\,T MPRAGE volumes ($0.7$\,mm isotropic, $256\times320\times320$). 32 subjects were used, 24 for the training partition and 8 for the testing partition, with no anatomy shared between them. Subjects were drawn from the \ac{hcp} subset for which 7\,T session data are also available, so that a future 7\,T release can reuse the same anatomies and field maps (Section~\ref{sec:future}). However, the simulations being released only use the 3\,T data.

The anatomical data were segmented with SPM8 \cite{Ashburner2005Unified} into six probabilistic tissue classes, in the following order: \ac{gm}, \ac{wm}, \ac{csf}, bone, soft tissue, and air. The classes for \ac{gm}, \ac{wm}, \ac{csf}, and soft tissue enter the forward model as the \emph{probabilistic tissue-fraction maps} $\Omega_{GM}$, $\Omega_{WM}$, $\Omega_{CSF}$, and $\Omega_{S}$. Each $\Omega_\tau(\mathbf{r}) \in [0,1]$ is the probability that voxel $\mathbf{r}$ belongs to tissue class $\tau$, and the maps are used as continuous partial-volume weights throughout, never thresholded into binary masks. This single term is used for them everywhere below. A brain mask was formed as $\Omega_{GM}+\Omega_{WM}+\Omega_{CSF} > 0.92$, followed by slice-wise hole filling.

The volume was truncated along the slice direction to a 128\,mm slab (183 source slices), and all volumes were resampled by cubic-spline interpolation to a $128\times128\times64$ high-resolution grid and, for the final data, to a $64\times64\times32$ grid. The resulting nominal voxel size of the released \ac{mrsi} matrix is $2.8\times3.5\times4.0$\,mm$^3$ over a $179.2\times224\times128$\,mm$^3$ field of view. 

Field maps were derived from the \ac{hcp} gradient-echo phase image with an echo-time difference of $\Delta\mathrm{TE}=2.46$\,ms and converted to Hz as $\Delta f = \phi/(4096 \cdot 2 \cdot 2.46) \times 10^{3}$, where $\phi$ is the stored phase. The resulting map was resampled to the simulation grid and the mean over the brain mask was subtracted so that $\Delta f$ is referenced to the mean water resonance of the brain. Within the brain mask, the released $B_0$ maps span roughly $-87$ to $+115$\,Hz at the 1st and 99th percentiles across subjects, with extreme values reaching $\pm 280$\,Hz near air--tissue interfaces (Table~\ref{tab:partitions}).

\subsection{Metabolite Signal Model}
\label{sec:metab}
Metabolite signals were generated from 16 quantum-mechanically simulated basis functions created with Vespa-Simulation (v1.1.1rc1, PyGAMMA backend) for a 3\,T system \cite{Soher2023Vespa}. The basis was simulated with an ideal one-pulse (90$^\circ$) sequence at $\mathrm{TE}=0$\,ms, a transmitter frequency of $123.24$\,MHz, $32{,}768$ points, and a sampling interval of $80.389$\,$\upmu$s (spectral width $12{,}439.5$\,Hz), with no $T_2$ decay applied at the simulation stage. Signals above $4.3$\,ppm were removed, and the resonance groups of tCr, PE, and Lac at or above $3.9$\,ppm were split into separate basis functions (tCr39, PE40, Lac41) because they are differentially affected by water suppression. The basis was then interpolated onto the acquisition raster and normalized to the tCr39 basis function (Appendix~\ref{app:implementation}). The 16 basis functions are listed in Table~\ref{tab:metabolites}.

At spatial location \(\mathbf{r}\), the metabolite signal was modeled as:
\begin{equation}
    \begin{split}
    s_M(\mathbf{r},t) = {}&
    W\!\left(z; T_1^{M}\right)
    e^{i2\pi \Delta f(\mathbf{r})t}\,
    g(\mathbf{r},t)\\
    &\times \left[
      \Omega_{GM}(\mathbf{r})\, s_{GM}(t)
    + \Omega_{WM}(\mathbf{r})\, s_{WM}(t)
    \right],
    \end{split}
    \label{eq:metabolite_signal}
\end{equation}
where \(\Delta f(\mathbf{r})\) is the local field offset, \(g(\mathbf{r},t)\) is the field-inhomogeneity lineshape term of Equation~\eqref{eq:voigt}, \(W(z;T_1^M)\) is the steady-state weighting of the slab-selective excitation given in Equation~\eqref{eq:ernst} below, and \(s_{GM}\) and \(s_{WM}\) are the tissue-specific metabolite signals
\begin{align}
    s_{GM}(t) &= \sum_{n=1}^{N_M} c_n^{GM}\,e^{-t/T_{2,n}^{GM}}\,\varphi_n(t),
    \label{eq:gm_metabolite_signal}\\
    s_{WM}(t) &= \sum_{n=1}^{N_M} c_n^{WM}\,e^{-t/T_{2,n}^{WM}}\,\varphi_n(t),
    \label{eq:wm_metabolite_signal}
\end{align}
with \(N_M=16\), \(\varphi_n(t)\) the \(n\)-th time-domain basis function, \(c_n^{GM}\) and \(c_n^{WM}\) the tissue-specific concentrations, and \(T_{2,n}^{GM}\) and \(T_{2,n}^{WM}\) the tissue-specific transverse relaxation times (Table~\ref{tab:metabolites}). Neither \(s_{GM}\) nor \(s_{WM}\) carries any spatial dependence: because the concentrations and relaxation times are fixed constants, the same two tissue signals are used in every voxel of every dataset in the release, and all spatial structure in \(s_M\) enters through the tissue fractions, the slab weighting, and the field map.

The factor \(W\) is the spoiled steady-state (Ernst) weighting of the slab-selective excitation \cite{Ernst1966FourierNMR},
\begin{equation}
    W(z;T_1) = \frac{1-e^{-\mathrm{TR}/T_1}}{1-\cos\theta(z)\,e^{-\mathrm{TR}/T_1}}\;\mathrm{Re}\left\{\sin\theta(z)\right\},
    \label{eq:ernst}
\end{equation}
with $\mathrm{TR}=450$\,ms. The slice-dependent flip angle enters through the Cayley--Klein parameters $\alpha(z),\,\beta(z)$ of a Shinnar--Le Roux-designed (\acs{slr}) slab-selective pulse \cite{Pauly1991SLR}, via $\cos\theta(z) = 1-2|\beta(z)|^2$ and $\sin\theta(z) = 2\,\alpha^{*}(z)\beta(z)$, designed for a $9.6$\,cm slab at a nominal flip angle of $40^\circ$ with a $1$\,ms pulse, and evaluated over the $12.8$\,cm extent of the 32 slices. A single relaxation time $T_1^{M} = 1300$\,ms was used for all metabolites in both tissue types. Because $W$ multiplies every metabolite identically, only the products $W\Omega_{GM}$ and $W\Omega_{WM}$ affect the released maps.

Metabolite concentrations and $T_2$ values were fixed constants, chosen to be consistent with published \textit{in vivo} \ac{gm} and \ac{wm} measurements pooled across \cite{Wang1998Differentiation,Inglese2008GlobalNAA,Safriel2005Reference,Hetherington1994Evaluation,Schott2010Short,Maudsley2009Mapping,McLean2000Quantitative,Soher1996Quantitative,Chard2002Brain}, and cross-checked against a random-effects meta-analysis of the study set compiled by \cite{Gudmundson2023Database}. No single reference set covers all 16 basis functions in both tissues at 3\,T, so the values adopted here were settled pragmatically rather than taken from one source, and they are reported in full in Table~\ref{tab:metabolites} so that any downstream result can be traced back to the exact numbers used. They enter Equations~\eqref{eq:gm_metabolite_signal} and \eqref{eq:wm_metabolite_signal} as the fixed constants of Table~\ref{tab:metabolites}, so the only per-subject variability in the metabolite term arises from $\Omega_{GM}$, $\Omega_{WM}$, and $\Delta f$.

\begin{figure*}
\centering
\includegraphics[width=1.0\textwidth,keepaspectratio]{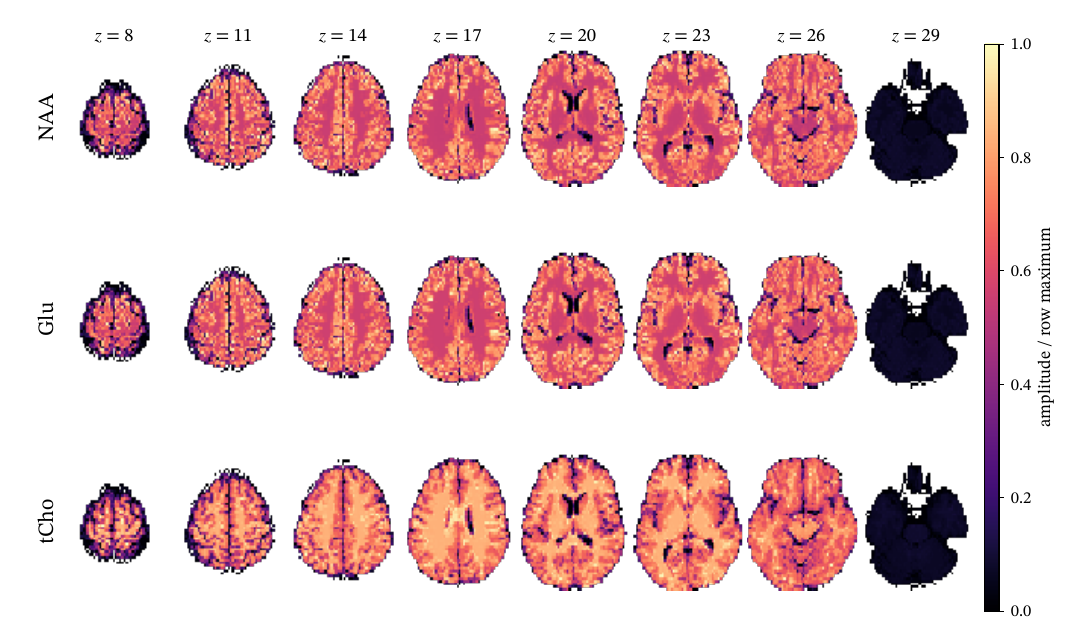}
\caption{NAA, Glu, and tCho axial slices through the excitation slab from training subject Sub4. NAA and Glu share the \ac{gm}-dominant pattern prescribed by their concentration vectors, whereas tCho, being \ac{wm}-dominant, inverts it. The near-uniform dimming at slice $z=29$ reflects the inferior roll-off of the slab profile $W(z;T_1^M)$ defined in Equation~\eqref{eq:ernst}, not a change in concentration: brain tissue fills that slice, but the peak amplitude has fallen to about $10\%$ of the plateau observed over $z=14$--$26$.
\label{fig:slabtrio}}
\end{figure*}

The effect of the slab weighting $W(z;T_1^M)$ is directly visible in the released ground truth (Figure~\ref{fig:slabtrio}): the metabolite amplitude is flat across the interior slices and falls away at the slab edges, uniformly for every metabolite, because $W$ multiplies them all identically. Any evaluation that scores metabolite maps slice by slice must account for this, since the amplitude drop at the slab edge is an excitation effect and not a concentration change.

Consequently, the ground-truth amplitude map of metabolite $n$ is exactly:
\begin{equation}
    a_n(\mathbf{r}) = W(z;T_1^M)\left[\Omega_{GM}(\mathbf{r})\,c_n^{GM} + \Omega_{WM}(\mathbf{r})\,c_n^{WM}\right],
    \label{eq:metamap}
\end{equation}
which is the quantity stored in the released \texttt{metaMap} array. Since $g(\mathbf{r},0)=1$, $a_n$ is the amplitude that the $n$-th basis function would have at $t=0$. The amplitude present in the first acquired sample is $a_n\,g(\mathbf{r},\mathrm{TE})$.

\begin{table*}
\caption{Metabolite basis functions and the fixed simulation parameters assigned to each basis function. Concentrations $c^{GM}$, $c^{WM}$ are given in mM. Values were chosen with reference to the meta-analysis compilation of Gudmundson et al.\ \cite{Gudmundson2023Database} but are not taken from it directly, see Section~\ref{sec:metab} for more details. Transverse relaxation times are given in ms. Lac and Lac41 are present in the basis but were assigned zero concentrations, so the corresponding ground-truth maps are identically zero. The suffixes 39, 40, and 41 denote the ppm of the resonance groups of tCr, PE, and Lac at or above $3.9$\,ppm, which were split off because they are affected by water suppression.
}\label{tab:metabolites}
\newcolumntype{C}{>{\centering\arraybackslash}X}
\begin{tabularx}{\textwidth}{lCCCC|lCCCC}
\toprule
\textbf{Basis} & \boldmath{$c^{GM}$} & \boldmath{$c^{WM}$} & \boldmath{$T_2^{GM}$} & \boldmath{$T_2^{WM}$} &
\textbf{Basis} & \boldmath{$c^{GM}$} & \boldmath{$c^{WM}$} & \boldmath{$T_2^{GM}$} & \boldmath{$T_2^{WM}$}\\
\midrule
Asp    & 3.70  & 2.07 & 200 & 200 & PE     & 2.12 & 2.12 & 200 & 200\\
GABA   & 1.86  & 0.88 & 200 & 200 & PE40   & 2.12 & 2.12 & 200 & 200\\
Gln    & 3.90  & 2.20 & 200 & 200 & Tau    & 1.93 & 1.74 & 200 & 200\\
Glu    & 8.60  & 6.00 & 200 & 200 & mIns   & 4.30 & 3.10 & 200 & 200\\
Lac    & 0     & 0    & 200 & 200 & sIno   & 0.50 & 0.57 & 200 & 200\\
Lac41  & 0     & 0    & 200 & 200 & tCho   & 1.38 & 1.78 & 235 & 239\\
NAA    & 10.98 & 7.49 & 261 & 351 & tCr    & 6.90 & 5.50 & 160 & 177\\
NAAG   & 1.40  & 2.60 & 261 & 351 & tCr39  & 6.90 & 5.50 & 160 & 177\\
\bottomrule
\end{tabularx}
\end{table*}

\subsection{Field-Inhomogeneity Lineshape}
Intra-voxel $B_0$ dispersion was modeled by an additional decay applied identically to all metabolites of a voxel,
\begin{equation}
    g(\mathbf{r},t) = \exp\left[-\frac{t}{T_2'(\mathbf{r})} - \frac{t^2}{T_G^2(\mathbf{r})}\right],
    \label{eq:voigt}
\end{equation}
which turns the Lorentzian lines produced by the $T_{2,n}$ terms into Voigt profiles. The Voigt profile is the lineshape generally observed \textit{in vivo}, arising from the combination of homogeneous (exponential) and inhomogeneous (Gaussian) broadening \cite{Marshall1997Voigt}. The exponential term $T_2^\prime$ accounts for the additional dephasing caused by field inhomogeneity ($\nicefrac{1}{T_2^\prime}=|\nabla B_0(\mathbf{r})|$), and $T_G$ contributes the Gaussian component. Both are derived from the subject $B_0$ map. Across the released data, $T_2^\prime$ has a median of $112$--$132$\,ms and its rate $1/T_2^\prime$ scales with the magnitude of the in-plane $B_0$ gradient, whereas $T_G$ is essentially a global constant, $114.0\pm2$\,ms (Appendix~\ref{app:implementation}). The Gaussian component therefore sets a fixed baseline linewidth, while the spatially varying part of the broadening is carried by $T_2^\prime$. The same term enters the macromolecular signal of Section~\ref{sec:mm}.

\subsection{Residual Water Signal Model}
\label{sec:water}
The unsuppressed water signal was represented as a tissue-weighted sum over \ac{gm}, \ac{wm}, \ac{csf}, and soft tissue:
\begin{equation}
    \begin{split}
    \rho_W(\mathbf{r},t) = {}&
    e^{i2\pi \Delta f(\mathbf{r})t}
    \sum_{\tau \in \{GM, WM, CSF, S\}}
    \Omega_{\tau}(\mathbf{r})\, c_W^{\tau}\\
    &\times W\!\left(z; T_{1,W}^{\tau}(\mathbf{r})\right) e^{-t/T_{2,W}^{\tau}(\mathbf{r})},
    \end{split}
    \label{eq:unsuppressed_water}
\end{equation}
where $c_W^{\tau}$ is the tissue water proton concentration, obtained from the water content of each tissue and the molarity of pure water (55.2\,M) \cite{deGraaf2019}. Unlike the metabolite term, the water relaxation times are perturbed independently in every voxel: $T_{2,W}^{\tau}(\mathbf{r})$ is drawn from a Gaussian with the tissue mean and a standard deviation of 5\,ms, and $T_{1,W}^{\tau}(\mathbf{r})$ is drawn uniformly from a small integer neighborhood (in ms) around the tissue mean. All water parameters are listed in Table~\ref{tab:water}.

Residual water after suppression was obtained from a Bloch-equation simulation of a \ac{wet} module with a single frequency-selective saturation pulse with a 50\,Hz bandwidth \cite{Ogg1994WET}. The simulation was tabulated once as a look-up table $M_z^{\mathrm{WET}}(z, \Delta f, T_1)$ over 64 slice positions, 1001 field offsets (1\,Hz steps, spanning approximately $\pm500$\,Hz), and 999 $T_1$ values. The suppressed signal replaces the steady-state weighting $W$ by the residual longitudinal magnetization:
\begin{equation}
    \begin{split}
    s_W(\mathbf{r},t) = {}&
    e^{i2\pi \Delta f(\mathbf{r})t}
    \sum_{\tau}
    \Omega_{\tau}(\mathbf{r})\, c_W^{\tau}\,
    M_z^{\mathrm{WET}}\!\left(z, \Delta f(\mathbf{r}), T_{1,W}^{\tau}(\mathbf{r})\right)\\
    &\times \mathrm{Re}\left\{\sin\theta(z)\right\}\,
    e^{-t/T_{2,W}^{\tau}(\mathbf{r})}.
    \end{split}
    \label{eq:suppressed_water}
\end{equation}

Because $M_z^{\mathrm{WET}}$ is indexed by $\Delta f(\mathbf{r})$, the effectiveness of water suppression varies spatially with the field map, which is the mechanism that produces spatially structured residual water in this dataset. Both the unsuppressed and the suppressed water volumes were generated on the high-resolution grid.

\begin{table}
\caption{Water signal simulation parameters. The proton concentration is calculated as the tissue water fraction multiplied by the molarity of pure water (55.2\,M). $T_1$ values are perturbed per voxel by a uniform integer offset $\pm\Delta T_1$ (in ms). $T_2$ values are perturbed per voxel by additive Gaussian noise with $\sigma = 5$\,ms.}\label{tab:water}
\newcolumntype{C}{>{\centering\arraybackslash}X}
\begin{tabularx}{\columnwidth}{lCCCCC}
\toprule
\textbf{Tissue} & \textbf{Water fraction} & \boldmath{$c_W$ (M)} & \boldmath{$T_1$ (ms)} & \boldmath{$\Delta T_1$ (ms)} & \boldmath{$T_2$ (ms)}\\
\midrule
\ac{gm}      & 0.80 & 44.16 & 1330 & $\pm2$  & $110 \pm 5$\\
\ac{wm}      & 0.70 & 38.64 &  830 & $\pm2$  & $80 \pm 5$\\
\ac{csf}     & 1.00 & 55.20 & 4500 & $\pm40$ & $1800 \pm 5$\\
Soft tissue  & 0.85 & 46.92 & 1200 & $\pm10$ & $60 \pm 5$\\
\bottomrule
\end{tabularx}
\end{table}

\subsection{Lipid Signal Model}
\label{sec:lipid}
The lipid contamination is not simulated parametrically. It is a measured signal consisting of a single high-resolution \textit{in vivo} \ac{fid}--\ac{mrsi} volume acquired at 3\,T from one donor subject and corrected for $B_0$ inhomogeneity. This volume was collected using an in-house \ac{fid}--\ac{epsi} sequence with a TR/TE of $300/4.7$\,ms, a transmitter frequency $123.198$\,MHz, and a dwell time $1.05$\,ms and a spectral readout of 251 points. The acquisition matrix utilized a matrix size of $128\times128\times64$ over a $240\times240\times192$\,mm$^3$ field of view, and the sequence included no water suppression, no outer-volume suppression, and no lipid-nulling module. Because the measurement is used directly, the lipid resonance amplitudes, frequencies, and decay constants are never made explicit anywhere in the forward model, and the lipid layer carries the line shapes, the resonance structure, and the spatial distribution of real subcutaneous fat rather than a sum of damped exponentials.

The lipid layer is applied to each simulated subject by non-rigid registration of the donor and subject $T_1$-weighted images with \texttt{antsRegistrationSyN} (rigid $+$ affine $+$ symmetric diffeomorphic \ac{syn}) \cite{Avants2008SyN}. The deformation field that maps donor space onto subject space is then applied, with linear interpolation, to every time point of the donor lipid \ac{fid} volume. It is that deformation field, and not a subject-specific lipid mask, which localizes the lipid layer because no subject-specific lipid masks were available. The donor volume was magnitude-thresholded and normalized by its spatial maximum before warping, and the registration settings are given in Appendix~\ref{app:implementation}. After registration, the subject-specific $B_0$ map was applied to introduce subject-dependent field inhomogeneity. Lipid signals were generated on the high-resolution grid, so their leakage into the brain is produced by the k-space truncation from Section~\ref{sec:truncation} rather than by an explicit spatial blur. The released data therefore contain unsuppressed lipids (Section~\ref{sec:future}).

Water and lipid signals are distributed together, summed into the single array \texttt{xtNuisance}, and are not released separately. They do, however, dominate different regions of the spectrum: residual water occupies the $4.0$--$5.4$\,ppm region and lipid the $0.6$--$1.9$\,ppm region. The amplitude of the nuisance signal integrated over each region reflects the total signal contributed by the respective simulation methods detailed in Sections~\ref{sec:water} and~\ref{sec:lipid}.

\begin{figure*}
\centering
\includegraphics[width=1.0\textwidth,keepaspectratio]{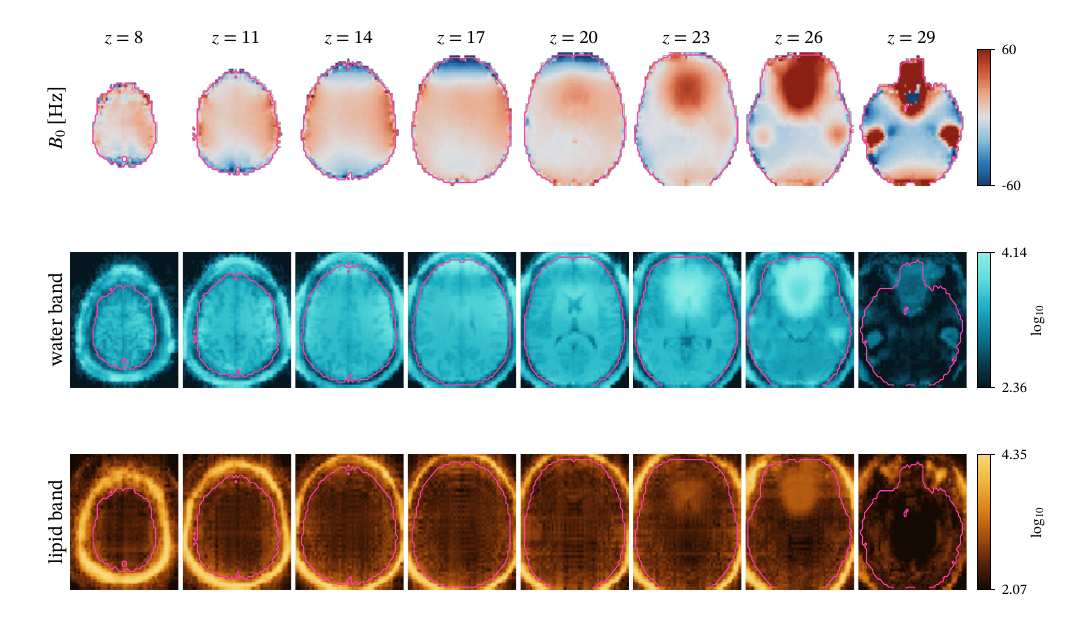}
\caption{The field map and the two nuisance bands across eight axial slices through the excitation slab in the testing dataset subject TestSub4. The brain mask is outlined in magenta, and each row shares one color scale. The water band is obtained by integrating the ground-truth nuisance signal over $4.0$--$5.4$\,ppm and the lipid band over $0.6$--$1.9$\,ppm. TestSub4 has the widest field-inhomogeneity range of the sub-challenge~1 testing data (Table~\ref{tab:partitions}). The extremes occur in the inferior slices, where the sinuses perturb the field. The lipid ring is present in every slice, with lipid leakage extending into the brain.}\label{fig:nuisslices}
\end{figure*}

Two mechanisms shape the nuisance signal. The residual water amplitude follows the frequency-selective profile of the \ac{wet} module of Equation~\eqref{eq:suppressed_water}. It is minimal on resonance and rises symmetrically with $|\Delta f|$, so that voxels at $|\Delta f| \approx 25$\,Hz retain roughly twice the residual water of voxels at $\Delta f = 0$. This is how the field map imprints itself on the residual water. The lipid signal, in contrast, is shaped by the k-space truncation of Section~\ref{sec:truncation}. Its amplitude falls by only about a factor of two over the first fifteen voxels of depth into the brain, given the slowly decaying tail of the sinc \ac{psf} of a rectangular window. Lipid contamination in this dataset is therefore non-local, and cannot be removed voxel by voxel. Figure~\ref{fig:nuisslices} shows how the field map and the two bands evolve through the slab.

\subsection{Macromolecular Signal Model}
\label{sec:mm}
Nine \textit{in vivo} macromolecular (MM) basis functions were derived similar to Pova\v{z}an et al.\ \cite{Povazan2018MM} to model broad background resonances in short-TE proton spectra \cite{Cudalbu2021MM,LaMaster2026SyntheticReview,Kreis2021Terminology}. 
These were acquired from six healthy volunteers at 7\,T using a 2D \ac{fid}-based double-inversion recovery \ac{mrsi} sequence. The sequence used a TR of 879\,ms, a TE$^*$ of 1.3\,ms, inversion times of TI$_1 = 570$\,ms and TI$_2 = 21$\,ms, and a flip angle of $55^\circ$. The acquisition used a $32\times32$ matrix over a $180\times180$,mm$^2$ field of view, with a nominal spatial resolution of $5.6\times5.6\times12$\,mm$^3$, a spectral readout of 2048 \ac{fid} points, and a 6\,kHz spectral bandwidth. Spectra were averaged across voxels and volunteers, and nine \ac{mm} components centered at 0.91, 1.21, 1.43, 1.67, 2.04, 2.26, 2.99, 3.21, and 3.77\,ppm were fitted with AMARES. The resulting components were frequency-scaled to 3\,T. \ac{gm}/\ac{wm} contrast was determined by fitting 7\,T \ac{mrsi} data from 10 volunteers with LCModel using a  basis set containing metabolites and \ac{mm}s. \ac{gm}, \ac{wm}, and \ac{csf} masks were generated with FAST from MPRAGE data, downsampled in k-space to match the \ac{mrsi} resolution, and thresholded with a tissue probability threshold $>0.7$. The mean \ac{mm} concentrations were calculated for each tissue fraction after applying quality control criteria. These criteria required a \ac{crlb} $<30\%$ for tCho, tNAA, and tCr, a \ac{crlb} $<900\%$ for the specific \ac{mm}, a linewidth $<25$\,Hz, and $\mathrm{SNR}>5$. Values were also excluded when the absolute \ac{crlb} exceeded half the mean concentration of the corresponding \ac{mm}. Additionally, the intrinsic \acs{fwhm} of each fitted component was verified against per-peak \textit{in vivo} estimates extrapolated to zero creatine linewidth, so that $g_B$ only needs to supply the additional voxel-wise field broadening. A tenth basis function, a creatine resonance at 3.9\,ppm damped over 190\,ms, serves solely as a normalization reference for the \ac{mm} amplitudes (Equation~\eqref{eq:mm_amp}). It is not injected into any simulated signal, but it is retained in the released \ac{mm} basis set \texttt{VtMM} (column~10) so that the normalization can be reproduced, and should be ignored when the basis set is used for fitting.

The macromolecular signal is:
\begin{equation}
    \begin{split}
    s_B(\mathbf{r},t) = {}&
    W\!\left(z;T_1^{MM}\right)
    e^{i2\pi \Delta f(\mathbf{r})t}\,
    g_B(\mathbf{r},t)\\
    &\times \sum_{n=1}^{9}
    \left[\Omega_{GM}(\mathbf{r})\,b_n^{GM} + \Omega_{WM}(\mathbf{r})\,b_n^{WM}\right]\varphi_{B,n}(t),
    \end{split}
    \label{eq:background_signal}
\end{equation}
with $T_1^{MM}=250$\,ms and amplitudes tied to the tissue-specific creatine concentration,
\begin{equation}
    b_n^{GM} = \kappa \, r_n^{GM}\, c_{\mathrm{tCr39}}^{GM}, \qquad
    b_n^{WM} = \kappa \, r_n^{WM}\, c_{\mathrm{tCr39}}^{WM},
    \label{eq:mm_amp}
\end{equation}
where $\kappa=0.4$ is a global \ac{mm} scaling factor. Ratios  $r_n^{GM}, r_n^{WM}$ (Table~\ref{tab:mm}) describe the concentration factor depending on the tissue type. For 1.21, 1.43, and 1.67\,ppm, the \ac{gm}/\ac{wm} contrast was unresolved, so $r_n^{GM}=r_n^{WM}$. The term $g_B$ models voxel-wise field-inhomogeneity broadening beyond each $\varphi_{B,n}$'s intrinsic \acs{fwhm}.

\begin{table}
\caption{Macromolecular basis functions defined by their resonance position and amplitude ratio relative to total creatine in \ac{gm} and \ac{wm}. The amplitudes enter the forward model as $b_n = \kappa\, r_n\, c_{\mathrm{tCr39}}$ with $\kappa=0.4$ (Equation~\eqref{eq:mm_amp}).}\label{tab:mm}
\newcolumntype{C}{>{\centering\arraybackslash}X}
\begin{tabularx}{\columnwidth}{lCCC}
\toprule
\textbf{Component} & \boldmath{$r_n^{GM}$} & \boldmath{$r_n^{WM}$} & \textbf{\ac{gm}/\ac{wm} ratio}\\
\midrule
MM 0.91\,ppm & 0.14049  & 0.14225  & 0.988\\
MM 1.21\,ppm & 0.24152  & 0.24152  & 1.000$^{\dagger}$\\
MM 1.43\,ppm & 0.27404  & 0.27404  & 1.000$^{\dagger}$\\
MM 1.67\,ppm & 0.24470  & 0.24470  & 1.000$^{\dagger}$\\
MM 2.04\,ppm & 0.099662 & 0.042158 & 2.364\\
MM 2.26\,ppm & 0.045433 & 0.020042 & 2.267\\
MM 2.99\,ppm & 0.20234  & 0.22049  & 0.918\\
MM 3.21\,ppm & 0.35101  & 0.39498  & 0.889\\
MM 3.77\,ppm & 0.33077  & 0.33268  & 0.994\\
\bottomrule
\end{tabularx}
\noindent{\footnotesize{$^{\dagger}$ \ac{gm} and \ac{wm} values were constrained to be equal because the measured contrast was not significant.}}
\end{table}

\subsection{Baseline and Noise Models}
\label{sec:baseline}
A smooth spectral baseline $s_{\mathrm{BL}}$ was added inside the brain mask, generated as a bounded, smoothed pseudo-random walk in the spectral domain \cite{LaMaster2025MRSSim}. The walk is drawn over $1$--$6$\,ppm at a density of 50 points per ppm, pinned to zero at both ends, taken with Gaussian steps of standard deviation $0.05$, and confined to $[-1,1]$. It is then smoothed with a window of $0.35$--$0.40$ of its length and scaled to between $0$ and $0.2$ of the spectral amplitude of the voxel. No points are dropped. The resulting baseline is identically zero outside the brain mask. Its time-domain envelope decays to $1\%$ of its peak within 19--20 sampling points ($\approx16$\,ms), so that it is smooth on the scale of the metabolite linewidths but not flat across the spectrum. The resulting baselines have a peak spectral amplitude of $12$--$17\%$ of the peak metabolite spectral amplitude in the same voxel, confirming the intended amplitude range. 
The baseline is drawn independently in each voxel rather than as a smooth spatial field, so methods that exploit spatial smoothness of the baseline will not find it here.

Noise was added as complex Gaussian noise, independent across voxels and time points, with equal variance in the real and imaginary channels and no spatial or temporal correlation. Estimated from voxels outside the brain mask, where all other components vanish, the per-channel standard deviation is $\sigma = 1.000\times10^{-3}$ in every training dataset. A restricted set of $\sigma \in \{0.600, 1.000, 1.400\}\times10^{-3}$ was used across the testing datasets (Table~\ref{tab:partitions}). 

\subsection{High-Resolution Synthesis and k-Space Truncation}
\label{sec:truncation}
A real \ac{mrsi} acquisition does not sample a continuous object. Instead, it samples a finite region of k-space and the image it returns is therefore the object convolved with the \ac{psf} of that sampling window. To reproduce this, the water and lipid components are first synthesized at higher spatial resolution than the released data and then reduced to the target matrix through k-space truncation, rather than by interpolation. Writing $y_{\mathrm{HR}}(\mathbf{r},t)$ for either component on the high-resolution grid, the low-resolution data are obtained by transforming to k-space, discarding the outer samples, and transforming back:
\begin{equation}
    \mathcal{T}\{y\}(\mathbf{r},t) = \mathcal{F}^{-1}_{\mathbf{k}}\Big[\mathcal{C}\left\{\mathcal{F}_{\mathbf{r}}\left[y_{\mathrm{HR}}(\mathbf{r},t)\right]\right\}\Big],
    \label{eq:kspace_truncation}
\end{equation}
where \(y\) is the component in question ($s_W$ or $s_L$), \(y_{\mathrm{HR}}\) is its high-resolution counterpart, \(\mathcal{F}_{\mathbf{r}}\) is the forward spatial Fourier transform taking the spatial coordinate \(\mathbf{r}\) to the spatial frequency \(\mathbf{k}\), \(\mathcal{F}^{-1}_{\mathbf{k}}\) is the inverse Fourier transform taking \(\mathbf{k}\) back to \(\mathbf{r}\) on the target grid, and \(\mathcal{C}\{\cdot\}\) is a rectangular cropping function that retains the central $64\times64\times32$ samples of k-space. Both transforms act on the three spatial dimensions only, the spectral dimension is untouched and the operator is applied independently at each of the 384 time points. No apodization or Hamming filter is applied, so the effective \ac{psf} is the sinc kernel of a rectangular k-space window. This step reproduces the partial-volume effects and the spatial bleeding of extracranial lipid signal into the brain that arise from the finite \ac{psf} of practical \ac{mrsi} acquisitions, whose severity is set by the encoded spatial resolution \cite{Motyka2019}. Explicit k-space simulation matters because spatial contamination in \ac{mrsi} is not equivalent to adding independent voxel-wise lipid or water signals \cite{LaMaster2026SyntheticReview}.

The metabolite, macromolecular, and baseline components were generated directly on the $64\times64\times32$ grid from spline-downsampled tissue maps and are therefore not subject to Equation~\eqref{eq:kspace_truncation}. The metabolite ground truth is consequently free of Gibbs ringing, while the nuisance signal is not.

\subsection{Data Format and Documentation}
The dataset is distributed in \ac{nifti} and \ac{nifti-mrs} formats \cite{Clarke2022NIFTIMRS}. These are the primary release: the anatomical image, the $B_0$ map, the brain mask, and each ground-truth metabolite map, named accordingly, are stored \ac{nifti} volumes while the spectroscopic volumes are \ac{nifti-mrs} files carrying the spectrometer frequency and dwell time in their header extension. Table~\ref{tab:nifti} explains the file naming convention per partition. Acquisition-like and grid parameters are summarized in Table~\ref{tab:acquisition}.

For convenience, each subject is additionally accompanied by a single MATLAB \texttt{.mat} file (HDF5, MATLAB v7.3) containing the same arrays together with the basis functions and the time vector. It is redundant with the \ac{nifti} release and is documented in Appendix~\ref{app:matfile}.

\begin{table*}
\caption{Files released per subject. All spectroscopic volumes are $64\times64\times32\times384$ complex-valued matrices. All image volumes are $64\times64\times32$, except the high-resolution anatomical reference, which is $128\times128\times64$. \texttt{XXX} denotes the 16 metabolite names listed in Table~\ref{tab:metabolites}, in lowercase.}\label{tab:nifti}
\newcolumntype{L}{>{\raggedright\arraybackslash}X}
\newcolumntype{C}{>{\centering\arraybackslash}p{0.95cm}}
\begin{tabularx}{\textwidth}{@{}lLCCC@{}}
\toprule
\textbf{File} & \textbf{Content} & \textbf{Train} & \textbf{Test} & \textbf{GT}\\
\midrule
\texttt{*\_mri\_t1w\_mpr.nii.gz}      & $T_1$-weighted image on the \ac{mrsi} grid & \checkmark & \checkmark & \\
\texttt{*\_mri\_t1w\_mpr\_HR.nii.gz}  & $T_1$-weighted image, high-resolution grid  &            & \checkmark & \\
\texttt{*\_mri\_b0\_map.nii.gz}       & Field offset $\Delta f$ in Hz               & \checkmark & \checkmark & \\
\texttt{*\_mri\_brain\_mask.nii.gz}   & Binary brain mask                           & \checkmark & \checkmark & \\
\midrule
\texttt{*\_mrs\_fids\_si\_data.nii.gz}     & Composite data $s$ (\ac{nifti-mrs})    & \checkmark & \checkmark & \\
\texttt{*\_mrs\_fids\_metabolites.nii.gz}  & Metabolite signal $s_M$ (\ac{nifti-mrs})& \checkmark &            & \\
\texttt{*\_mrs\_fids\_nuisance.nii.gz}     & Water $+$ lipid signal (\ac{nifti-mrs}) & \checkmark &            & \\
\texttt{*\_mrs\_metab\_result\_XXX.nii.gz} & Ground-truth amplitude map $a_n$        & \checkmark &            & \\
\midrule
\texttt{*\_all.mat}, \texttt{*\_all\_truth.mat} & All of the above, plus the basis functions and the time vector. The testing ground truth includes the separate \ac{mm} and baseline signals & \checkmark & \checkmark & \checkmark\\
\bottomrule
\end{tabularx}
\end{table*}

\begin{figure*}
\centering
\includegraphics[width=1.7\columnwidth,keepaspectratio]{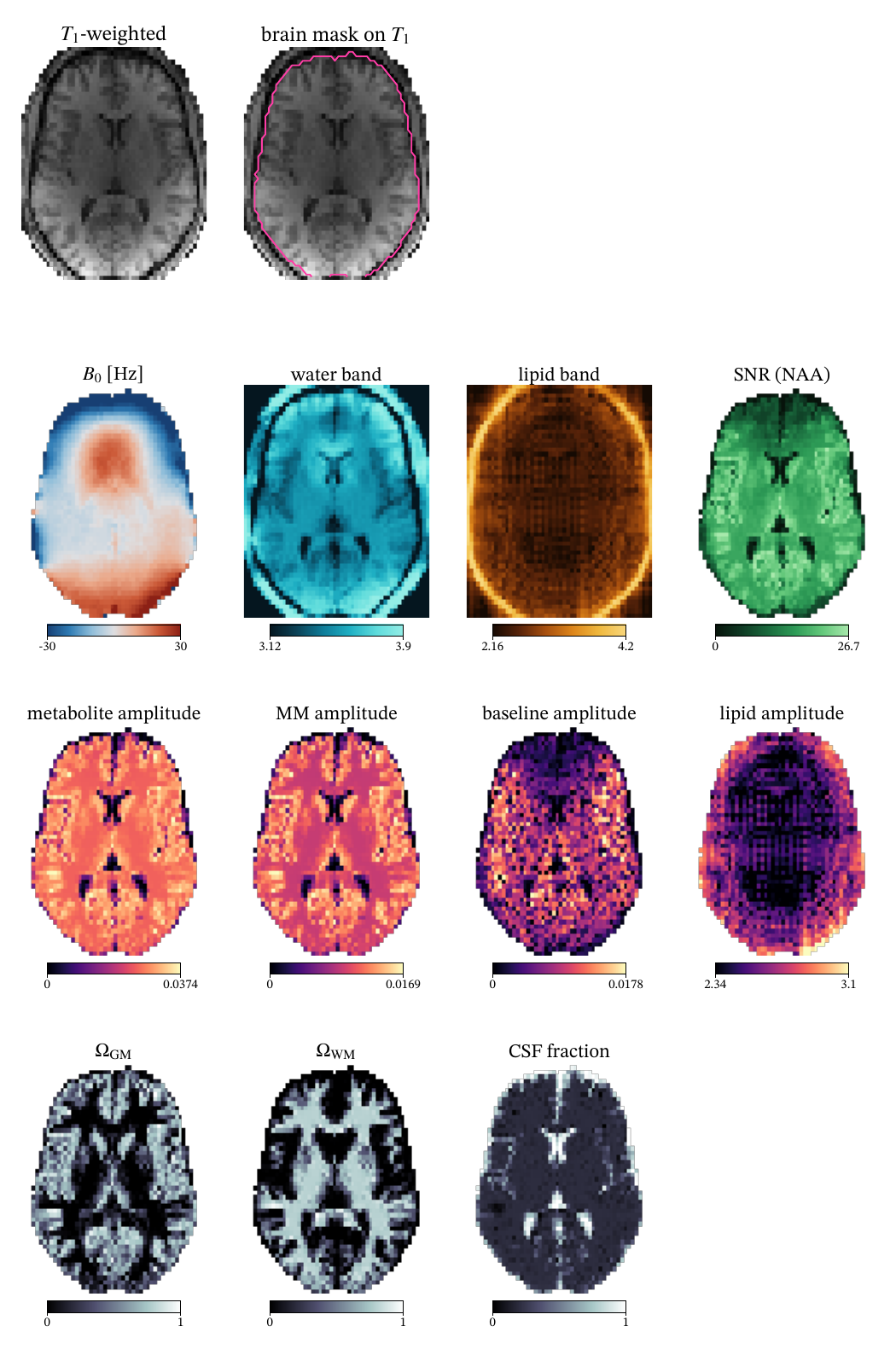}
\caption{Anatomy, field, tissue, nuisance, and quality maps for axial slice $z=20$ of 32 from the testing dataset subject TestSub1. Top row: the $T_1$-weighted reference with the brain mask outlined in magenta. Second row: the $B_0$ map, the residual-water and lipid band amplitudes, each corresponding to the ground-truth nuisance signal integrated over its spectral region, and the per-voxel \ac{snr} defined as in Table~\ref{tab:partitions}. Third row: the amplitudes of the metabolite, macromolecular, baseline, and lipid components. And in the fourth row: the \ac{gm} and \ac{wm} fractions are recovered by solving Equation~\eqref{eq:metamap}, together with the residual \ac{csf} fraction $1-\Omega_{GM}-\Omega_{WM}$.}
\label{fig:anatomy}
\end{figure*}

Figure~\ref{fig:anatomy} collects, for one slice, every map that enters or emerges from the forward model. Two features of that set are not obvious from the equations. The nuisance amplitude in the lipid region is confined to the subcutaneous ring, while that in the water region is not. The baseline amplitudes, alone among the simulation components, carry no anatomical structure whatsoever.

\begin{table*}
\caption{Acquisition-like and simulation parameters for the released datasets.}\label{tab:acquisition}
\newcolumntype{L}{>{\raggedright\arraybackslash}X}
\begin{tabularx}{\textwidth}{lL}
\toprule
\textbf{Parameter} & \textbf{Value}\\
\midrule
Field strength / transmitter frequency & 3\,T / $123.24$\,MHz ($1$H)\\
Nominal sequence & \ac{fid}--\ac{epsi}, slab-selective \acs{slr} \\
\ac{tr} / \ac{te} & $450$\,ms / $1.66$\,ms\\excitation
Dwell time $\delta t$ / spectral width & $0.83$\,ms / $1204.8$\,Hz\\
Sampling grid & $t_n = (n+2)\delta t$, $n=0,\dots,383$\\
Number of \ac{fid} points & 384\\
Final matrix & $64\times64\times32\times384$\\
High-resolution synthesis matrix (water, lipid) & $128\times128\times64\times384$\\
Nominal voxel size & $2.8\times3.5\times4.0$\,mm$^3$ (derived, see text)\\
Field of view & $179.2\times224\times128$\,mm$^3$ (derived, see text)\\
Excitation slab thickness & $96$\,mm (nominal flip angle $40^\circ$)\\
Metabolite basis & 16 functions, Vespa v1.1.1rc1, one-pulse, $\mathrm{TE}=0$\\
Basis raster (as distributed) & $32{,}768$ points, $80.389$\,$\upmu$s, $12{,}439.5$\,Hz\\
\ac{mm} basis & 9 \textit{in vivo} \ac{hlsvd} components ($+1$ tCr reference)\\
Water suppression & \ac{wet}, one frequency-selective pulse, 50\,Hz bandwidth\\
Lipid & measured \textit{in vivo}, non-rigidly registered, unsuppressed\\
$k$-space filter & rectangular crop, no apodization\\
Noise & complex Gaussian, i.i.d., $\sigma$ per channel (Table~\ref{tab:partitions})\\
\bottomrule
\end{tabularx}
\end{table*}

\section{Results}
\subsection{Dataset Contents}
The released dataset comprises 32 subject-level datasets: 24 training (\texttt{Sub1}--\texttt{Sub24}) and 8 testing. Five testing datasets belong to sub-challenge~1 (\texttt{TestSub1}--\texttt{TestSub5}) and carry the nuisance ground truth \texttt{xtNuisance}. The remaining three belong to sub-challenge~2 (\texttt{TestSub10}--\texttt{TestSub12}), in which no water or lipid signal was injected, so there is no nuisance array. Training and testing subjects are drawn from disjoint sets of \ac{hcp} subjects, so there is no anatomy shared between the partitions. Each dataset provides contaminated \ac{fid}--\ac{mrsi} data, a $T_1$-weighted anatomical image, a \(B_0\) map, a brain mask, and simulation metadata. There are ground-truth metabolite maps and component signals for both partitions. The macromolecular and baseline components are only distributed with the testing ground truth (Table~\ref{tab:nifti}).

\begin{figure*}
\centering
\includegraphics[width=1.0\textwidth,keepaspectratio]{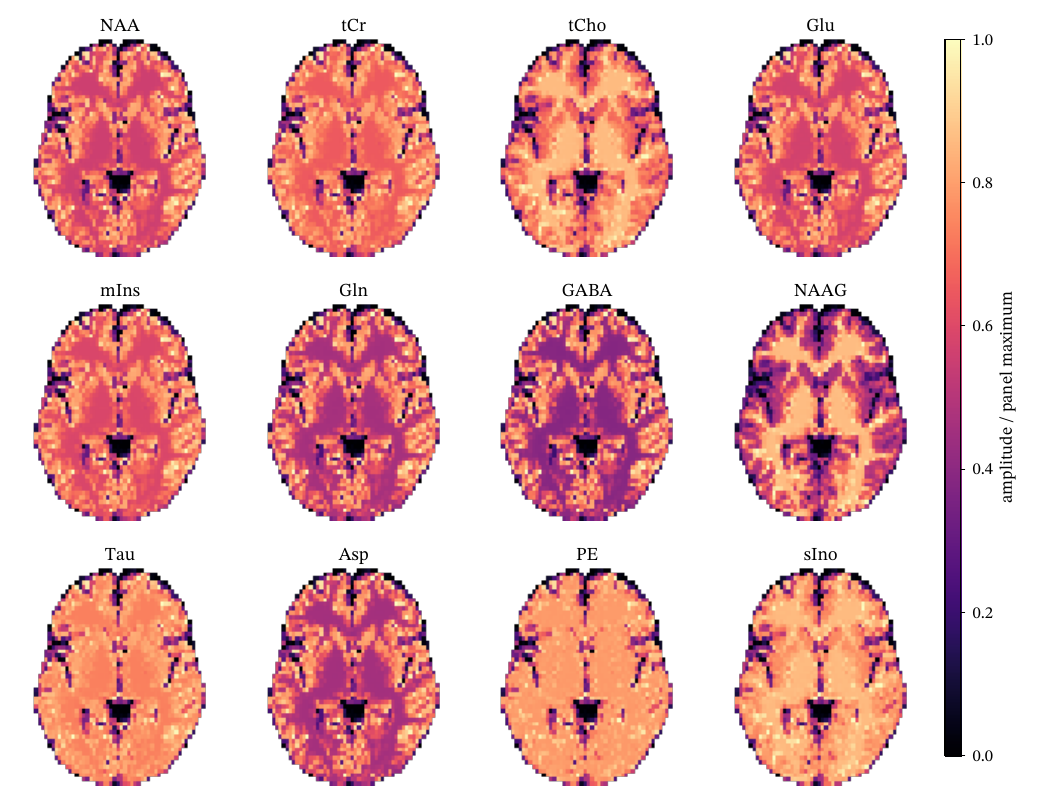}
\caption{Ground-truth metabolite amplitude maps for axial slice $z=22$ of 32 from training subject Sub13. The 12 metabolites assigned nonzero concentrations are shown.}
\label{fig:metabmaps}
\end{figure*}

The ground-truth amplitude maps are shown in Figure~\ref{fig:metabmaps}. NAA, tCr, Glu, mIns, Gln, GABA and Asp are \ac{gm}-dominant while tCho, NAAG and sIno are \ac{wm}-dominant. Lac and Lac41 were assigned zero concentration, and their maps are identically zero. By Equation~\eqref{eq:metamap} each map is a linear combination of $\Omega_{GM}$ and $\Omega_{WM}$ weighted by $W(z;T_1^M)$. The twelve maps with non-zero concentration span a two-dimensional subspace, and carry two independent images between them. This holds for every dataset in the release.

\subsection{Example Maps and Spectra}
\label{sec:results_spectra}
\begin{figure*}
\centering
\includegraphics[width=0.7\textwidth,keepaspectratio]{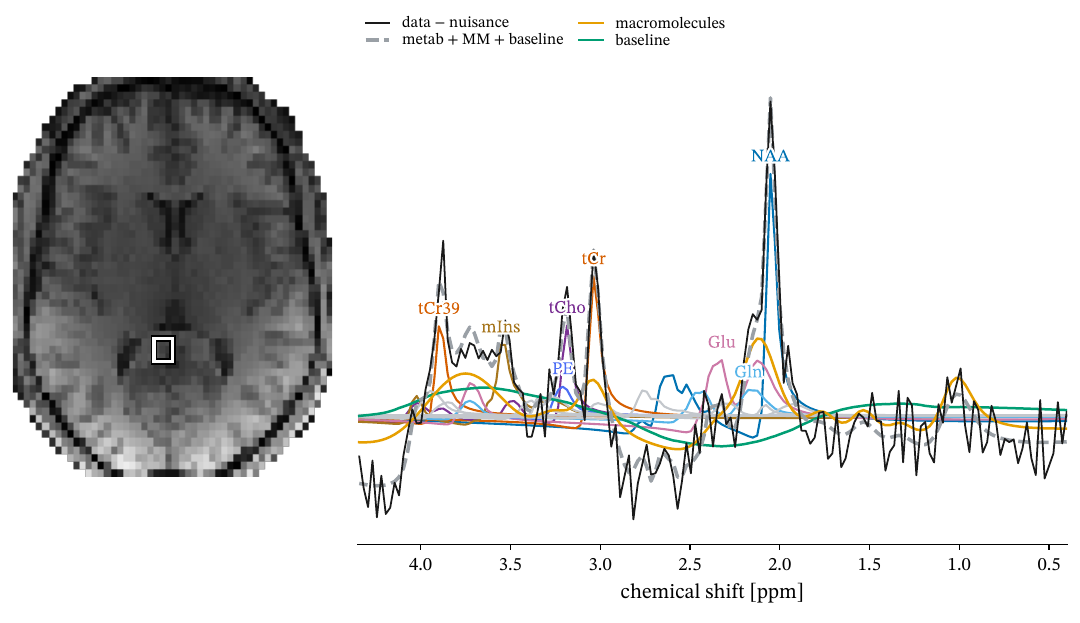}
\caption{Ground-truth signal components for a \ac{gm}-rich, well-shimmed voxel with $\Delta f \approx 0$\,Hz in the axial slice $z=20$ from subject TestSub1. The voxel is marked on the reference image. Black: the acquired data after subtraction of the ground-truth nuisance signal, i.e.,\ metabolites $+$ \ac{mm} $+$ baseline $+$ noise. Dashed gray: the sum of the three noiseless ground-truth components, which differs from the black curve only by the noise. Orange and green: the macromolecular and baseline components, respectively. The metabolite term, colored and labeled, is shown as individual basis contributions. The 8 largest are labeled by named, and the remainder are shown in light gray.}
\label{fig:components}
\end{figure*}

The simulation components are stored separately, and their sum reproduces the composite data. Figure~\ref{fig:components} shows the decomposition for one \ac{gm}-rich voxel, with the metabolite term broken down into the contribution of each basis function: $s_M+s_B+s_{\mathrm{BL}}$ tracks the acquired, nuisance-subtracted data to within the noise. At $\mathrm{TE}=1.66$\,ms, the macromolecular background reaches a substantial fraction of the metabolite amplitude and overlaps the NAA, tCr, and tCho resonances directly. The baseline is smooth on the scale of the metabolite linewidths and comparable in amplitude to the smaller metabolites.

\begin{figure*}
\centering
\includegraphics[width=0.85\textwidth,keepaspectratio]{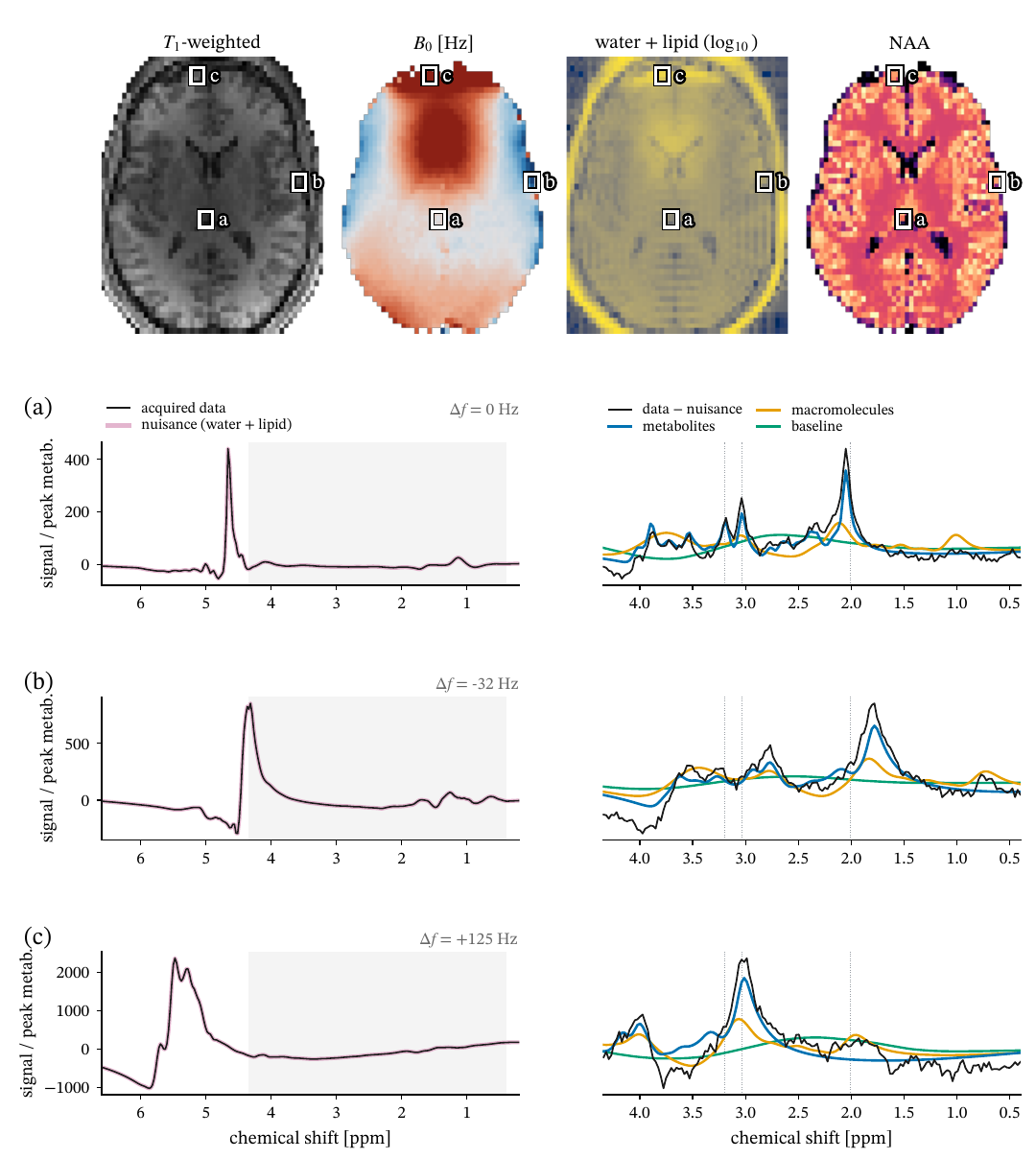}
\caption{Maps and spectra from the axial slice $z=20$ of 32 for testing dataset subject TestSub2. Top row, from left to right: the $T_1$-weighted reference, the $B_0$ map, the nuisance amplitude, and the ground-truth NAA map, each of which with three voxels marked. The voxels span the conditions the dataset is designed to test. Their array indices $(x,y,z)$ are given so that each spectrum can be reproduced directly from the release: (a) a well-shimmed deep-brain voxel at $(36,33,20)$ with $\Delta f = 0$\,Hz; (b) a cortical-rim voxel at $(9,35,20)$ adjacent to the lipid ring with $\Delta f = +125$\,Hz; and (c) a frontal voxel at $(29,11,20)$ near the sinus with $\Delta f = -32$\,Hz. Bottom rows: for each voxel marked in the top row, the acquired data with the ground-truth nuisance overlaid (left, the shaded band marks the region shown on the right), and the same data after nuisance subtraction, with the ground-truth metabolite, macromolecular, and baseline components (right). TestSub2 uses $1.4\times$ the training noise level (Table~\ref{tab:partitions}). Voxel (c) shows a displacement of the whole spectrum of nearly $1$\,ppm across the entire spectrum.}\label{fig:composite}
\end{figure*}

Figure~\ref{fig:composite} places the same decomposition beside the maps that locate it, for three voxels from TestSub2 spanning a range of $B_0$ field offsets and lipid proximities. Across the training partition, the peak nuisance amplitude exceeds the peak metabolite amplitude by a factor of $2000$--$2950$. At voxel~(c), where $\Delta f = +125$\,Hz, the metabolite spectrum is displaced by almost one ppm.

\begin{figure*}
\centering
\includegraphics[width=0.85\textwidth,keepaspectratio]{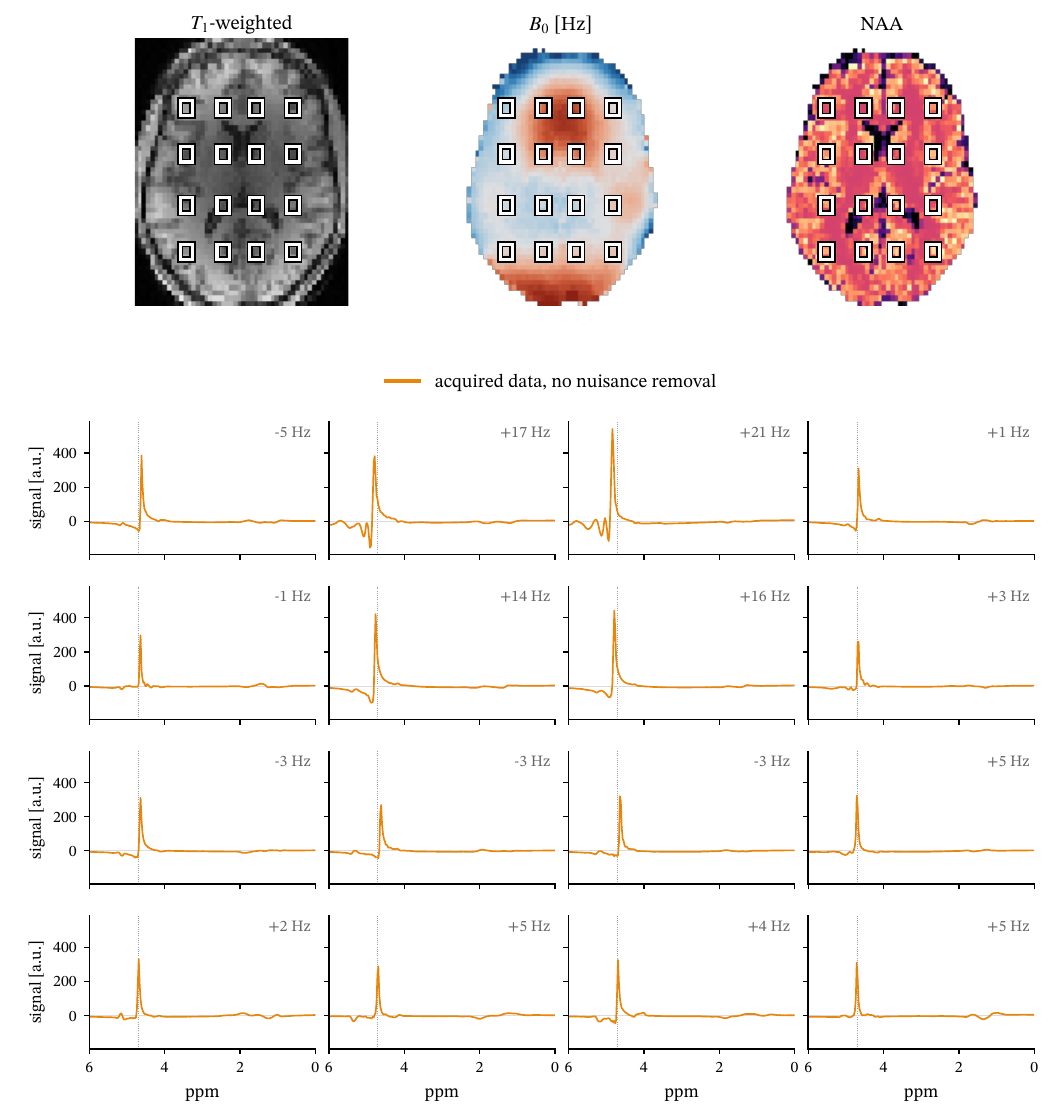}
\caption{The acquired data for 16 voxels sampled on a lattice in axial slice $z=20$ of 32 from testing dataset subject TestSub3, shown over $0$--$6$\,ppm without nuisance removal. Top row: the $T_1$-weighted reference, the $B_0$ map, and the ground-truth NAA map, with the 16 sampled voxels marked. Bottom rows: for each voxel, the acquired data, with the local field offset is annotated. All 16 plots share one vertical scale, so the residual water amplitude can be compared directly between voxels, which varies with the local field offset because of the frequency selectivity of the \ac{wet} module. The dotted line marks the water resonance at $4.70$\,ppm. The same voxels are shown after nuisance subtraction, over the fitting window, in Figure~\ref{fig:spectragrid}.}
\label{fig:rawgrid}
\end{figure*}

Figure~\ref{fig:rawgrid} shows the acquired data for sixteen voxels on a lattice of one subject from the testing dataset, over the full spectral range. The residual water resonance at $4.70$\,ppm exceeds the peak metabolite amplitude by a median factor of $1.1\times10^{3}$ over brain voxels. Its amplitude varies between voxels with the local field offset. On this scale, the metabolite spectrum is not distinguishable.

\begin{figure*}
\centering
\includegraphics[width=0.85\textwidth,keepaspectratio]{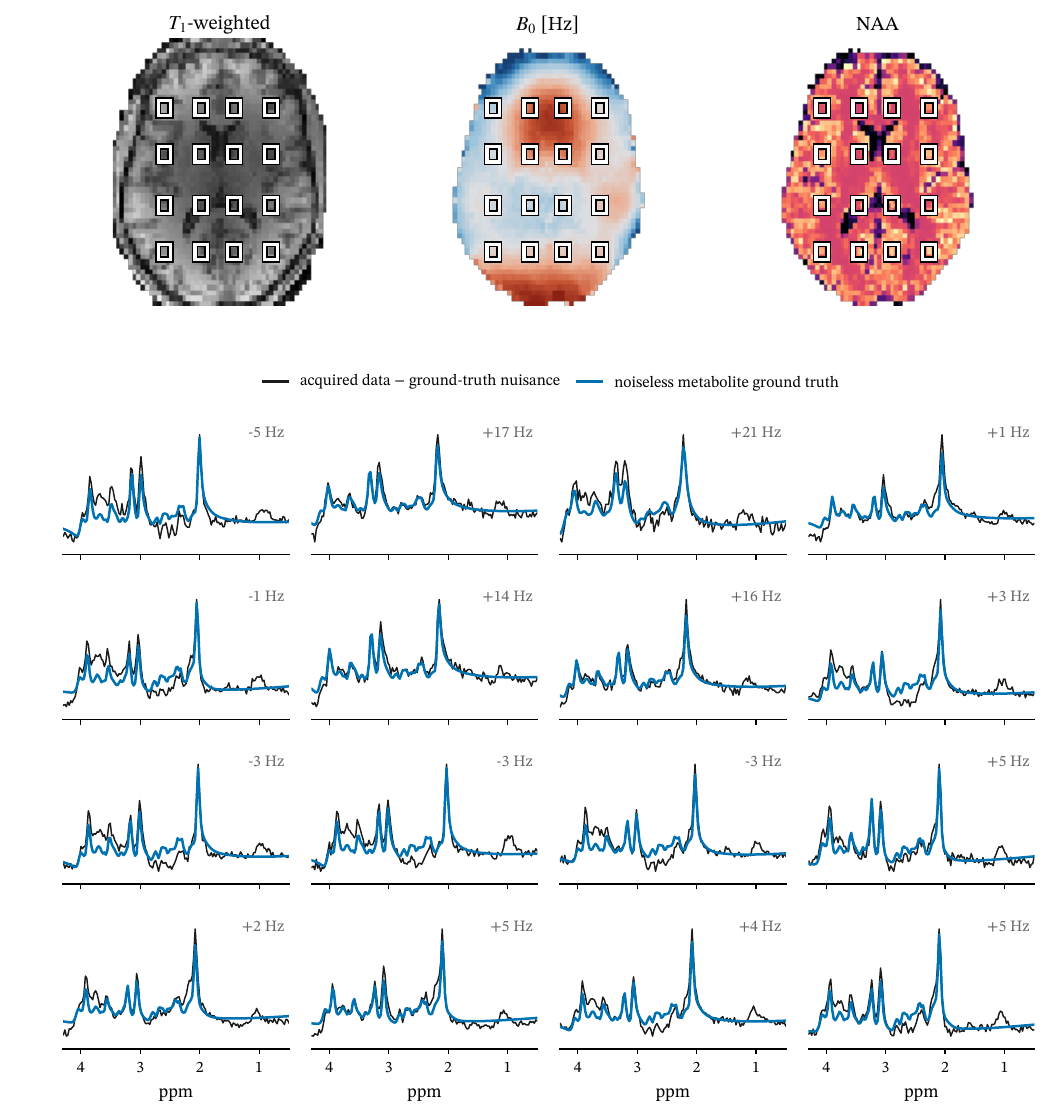}
\caption{The 16 voxels of Figure~\ref{fig:rawgrid} after nuisance subtraction for the axial slice $z=20$ of 32 from testing dataset subject TestSub3. Top row: the $T_1$-weighted reference, the $B_0$ map, and the ground-truth NAA map, with the 16 sampled voxels marked. Bottom rows: for each voxel, the acquired data after subtraction of the ground-truth nuisance signal (black) and the noiseless metabolite ground truth (blue), with the local field offsets annotated. The difference between the black and blue curves comprises the macromolecular signal, baseline, and noise. The progressive spectral shift with $\Delta f$ with the spectral shape preserved, is the signature of the frequency modulation in Equation~\eqref{eq:metabolite_signal}.}
\label{fig:spectragrid}
\end{figure*}

Figure~\ref{fig:spectragrid} shows the same sixteen voxels after subtraction of the ground-truth nuisance signal, over the $0.4$--$4.35$\,ppm window. That window excludes the water resonance, but the water wings and the lipid signal within it still exceed the peak metabolite amplitude by a median factor of $12$. The noiseless metabolite ground truth tracks the nuisance-subtracted data in every voxel, and the spectra are progressively frequency shifted by $\Delta f$ across the slice, according to the local $B_0$ field offset. 

Table~\ref{tab:partitions} reports the noise level, spectral \ac{snr}, and field-inhomogeneity range measured from every released dataset. All \ac{snr} values are quoted for the unapodized spectrum against an intrinsic NAA linewidth of $\approx16$\,Hz. Applying 3\,Hz of exponential broadening reduces the spectral noise standard deviation by a factor of $2.4$ and approximately doubles the apparent \ac{snr}.

Within the training partition, the per-channel noise level is $\sigma = 1.000\times10^{-3}$ in every dataset and the median \ac{snr} spans $16.2$--$17.8$ across the 24 subjects. The metabolite concentrations are identical across subjects, so the inter-subject variability that remains is anatomical: tissue fractions, brain coverage, and the $B_0$ field map.

The testing partition differs from the training partition in a small number of controlled ways. TestSub2 and TestSub11 were generated with $1.4\times$ the training noise level and TestSub3 and TestSub12 with $0.6\times$. The median \ac{snr} values scale accordingly. TestSub5 has a reduced nuisance amplitude and the largest baseline-to-metabolite ratio. The widest field-inhomogeneity ranges occur in TestSub4 and TestSub10. The $B_0$ maps derive from subject anatomy, so these ranges are measured rather than imposed.

\begin{table*}
\caption{Parameters measured from all 32 released datasets. $\sigma$ is the per-channel noise standard deviation, estimated from voxels outside the brain mask where all other components vanish. \ac{snr} is defined as the peak magnitude of the noiseless metabolite spectrum in the 1.5--4.3\,ppm range divided by the standard deviation of the spectral noise $\sigma\sqrt{N_t}$. This quantity is evaluated per voxel over three representative slices. The median and the 5th--95th percentile range across brain voxels are reported. The $\Delta f$ column gives the 1st and 99th percentiles over brain voxels of the whole volume. For the training partition, it gives the smallest 1st percentile and the largest 99th percentile across the 24 subjects. Low 5th-percentile \ac{snr} values arise in partial-volume voxels at the edge of the excitation slab.}\label{tab:partitions}
\newcolumntype{C}{>{\centering\arraybackslash}X}
\begin{tabularx}{\textwidth}{lcCCCcc}
\toprule
\textbf{Dataset} & \textbf{Sub-ch.} & \boldmath{$\sigma\,(10^{-3})$} & \textbf{\ac{snr} median} & \textbf{\ac{snr} 5--95\%} & \boldmath{$\Delta f$ (Hz)} & \textbf{Nuisance}\\
\midrule
Sub1--Sub24  & 1, 2 & $1.000$ & $16.2$--$17.8$ & $0.0$--$24.8$ & $-87$ to $115$ & yes\\
\midrule
TestSub1  & 1 & $1.000$ & $16.6$ & $1.7$--$23.5$ & $-62$ to $85$   & yes\\
TestSub2  & 1 & $1.400$ & $12.4$ & $0.9$--$17.4$ & $-81$ to $93$   & yes\\
TestSub3  & 1 & $0.600$ & $27.9$ & $2.9$--$39.5$ & $-69$ to $97$   & yes\\
TestSub4  & 1 & $1.000$ & $17.0$ & $0.0$--$24.0$ & $-62$ to $116$  & yes\\
TestSub5  & 1 & $1.000$ & $17.0$ & $0.4$--$24.6$ & $-57$ to $81$   & reduced\\
\midrule
TestSub10 & 2 & $1.000$ & $15.6$ & $1.8$--$23.6$ & $-113$ to $109$ & no\\
TestSub11 & 2 & $1.400$ & $12.1$ & $0.1$--$17.3$ & $-67$ to $102$  & no\\
TestSub12 & 2 & $0.600$ & $28.1$ & $2.4$--$39.2$ & $-69$ to $93$   & no\\
\bottomrule
\end{tabularx}
\end{table*}

\subsection{Benchmark Use Case}
\label{sec:benchmark}
The dataset was used as the basis for a community processing and quantification benchmark with two sub-challenges. Sub-challenge~1 required nuisance-signal removal followed by spectral quantification on the five testing datasets (TestSub1--5), for which the nuisance ground truth \texttt{xtNuisance} is supplied. Sub-challenge~2 required spectral quantification alone, on the three testing datasets (TestSub10--12), into which no water or lipid signal was injected. Groups could enter either or both challenges. The ground truth for the testing partitions was withheld from participants and used by the organizers for evaluation. Because the ground truth is available per component and per metabolite, and because the $B_0$ field map, the tissue fractions, and the excitation-slab profile are all known, error can be resolved by metabolite, by tissue class, by slice within the excitation slab, and by $B_0$ field offset.

\section{Discussion}
The 2024 \ac{mrsi} Data Processing and Quantification Challenge Synthetic Dataset fills a practical gap between idealized spectrum simulation and \textit{in vivo}-only validation. It supplies known ground-truth metabolite maps while retaining the sources of spatial and spectral difficulty that make \ac{mrsi} processing hard: subject-specific anatomy and tissue fractions, \(B_0\)-dependent frequency and lineshape variation, residual water, lipid contamination and realistic voxel bleeding through a finite point-spread function, a spectral baseline, and noise. Several of the failure modes it exposes arise only where spectral and spatial effects interact and are invisible to single-voxel simulations.

The measurements of Section~\ref{sec:results_spectra} illustrate how demanding this regime is. The nuisance signal dominates, with its peak amplitude exceeding the peak metabolite amplitude by a factor of $2000$--$2950$ across the training partition. As a result, the entire metabolite spectrum sits within the linewidth of the residual water resonance. Excluding that resonance does not resolve the situation, because the water wings and the lipid signal still exceed the peak metabolite amplitude by a median factor of $12$ inside the $0.4$--$4.35$\,ppm fitting window. Metabolite quantification therefore requires recovering a signal that is consistently and substantially weaker than the overlying nuisance signal. The lipid contamination, moreover, is not local. Its amplitude decreases by only about a factor of two over the first fifteen voxels of depth into the brain, following the slowly decaying tail of the sinc \ac{psf} of the rectangular k-space window (Section~\ref{sec:lipid}). Therefore, it cannot be removed voxel by voxel. Nuisance removal is thus as much a spatial problem as a spectral one, which distinguishes this dataset from a stack of independent single-voxel spectra.

Two additional properties challenge some common simplifying assumptions. At $\mathrm{TE}=1.66$\,ms, the macromolecular background signal reaches a substantial fraction of the metabolite amplitude and overlaps the NAA, tCr and tCho resonances directly (Figure~\ref{fig:components}). This is not a slowly varying term that a flexible baseline can absorb. And the field offsets span roughly $-113$ to $+116$\,Hz across the release (Table~\ref{tab:partitions}), displacing whole spectra by close to one ppm (Figure~\ref{fig:composite}). A fitting prior centred on the nominal peak positions fails at that displacement.

The dataset also aligns with current synthetic \ac{mrs} guidance. The ISMRM \ac{mrs} Synthetic Data Working Group review asks that synthetic data be tailored to the intended application, that assumptions be reported transparently, that variability be realistic, and that formats and metadata be accessible \cite{LaMaster2026SyntheticReview}. The proposed \ac{mrssynmrs} table extends the \ac{mrsinmrs} minimum reporting standards for \textit{in vivo} \ac{mrs} \cite{Lin2021MRSinMRS} to synthetic data. It asks specifically for the simulation details that determine whether a dataset can be reused for validation or training. Appendix~\ref{app:checklist} completes that table here. Where the generating code was not archived, the entry reports the value measured from the released arrays.

\subsection{Implications for Evaluation}
\label{sec:evaluation}
This dataset follows two precedents. The ISMRM \ac{mrs} fitting challenge established synthetic data with known ground truth as the basis for comparing quantification software \cite{Marjanska2022Challenge}. The 2023 ISBI challenge to reduce GABA-edited \ac{mrs} acquisition time scored machine-learning reconstruction pipelines against a known ground truth in its simulated track \cite{Berto2024GABAChallenge}. The challenge then ranked the submissions on mean squared error, \ac{snr}, linewidth, and a shape score together rather than on any single metric, because no one of them captures reconstruction quality alone.

The same logic applies here, and the structure of the release makes it actionable. Ground truth exists per component and per metabolite, and the $B_0$ field map, the tissue fractions, and the excitation-slab profile are known for every dataset (Section~\ref{sec:benchmark}). Error can therefore be resolved by metabolite, by tissue class, by slice within the excitation slab, and by $B_0$ field offset. A single global error over the whole volume conceals precisely the behavior the dataset was built to expose.

The excitation slab needs particular care. The metabolite amplitude is flat across the interior slices and falls away at the slab edges. It does so uniformly for every metabolite, because $W(z;T_1^M)$ multiplies them all identically. At $z=29$, it has fallen to about $10\%$ of the plateau reached at $z=14$--$26$ (Figure~\ref{fig:slabtrio}). Slice-wise scoring that reads that drop as a concentration change is measuring the excitation profile rather than the actual metabolic profile.

Two properties of the training partition bear on this directly. Its \ac{snr} is homogeneous, with a median $16.2$--$17.8$ across the 24 subjects, at a fixed noise level of $\sigma=1.000\times10^{-3}$ (Table~\ref{tab:partitions}). Any method tuned on it has seen essentially one noise condition. And the metabolite concentrations are fixed constants, so the twelve non-zero ground-truth maps carry two independent images between them: all spatial structure originates in the tissue segmentation and in the excitation slab, meaning that none of it is metabolic.

\subsection{Limitations and Appropriate Use}
\label{sec:limitations}
Three properties of the release warrant highlighting. \texttt{xtMeta} contains only the metabolite signal. The macromolecular component is present in the composite data, i.e., \texttt{xtAll}, but is only distributed separately for the testing ground truth. The appropriate algorithm input for sub-challenge 2 is therefore \texttt{xtAll}$-$\texttt{xtNuisance}, with \texttt{xtMeta} or \texttt{metaMap} as the target. Consensus preprocessing and quantification practices are laid out in \cite{Near2020Consensus}. The ground-truth maps are defined on the target grid and are exempt from the k-space truncation of Section~\ref{sec:truncation}, which acts only on the water and lipid signal, so an algorithm operating on the truncated data cannot recover them exactly, even in principle. Lastly, spline downsampling of the tissue-probability maps tends to undershoot boundaries, leaving small negative values in \texttt{metaMap}, as low as $-3.1\times10^{-4}$, or about 5\% of the peak NAA amplitude, in roughly 8\% of voxels. Metrics that assume a non-negative ground truth require those voxels to be masked or clipped.

For ML and DL work, the dataset is useful within a boundary worth stating precisely. The metabolite concentrations and relaxation times are fixed constants shared by all 32 datasets, so the only variability in the metabolite ground truth is anatomical, and the ground-truth maps span a two-dimensional subspace. A model trained here can learn the concentration vectors themselves, and its error on the testing partition can then overstate its performance on data carrying genuine biological variability. The noise level is likewise constant within the training partition. The dataset therefore suits controlled stress testing and the comparison of processing algorithms. It is not a substitute for training data with realistic biological variability.

The simulation does not reproduce all \textit{in vivo} variability. Subject motion \cite{Andronesi2021Motion}, physiological instability, eddy currents, gradient imperfections, coil-sensitivity variation, imperfect excitation and refocusing pulses, scanner-vendor differences, non-Lorentzian line shapes, and pathology-specific metabolic changes are absent or simplified. The concentration and relaxation assumptions are bounded by the available literature, particularly for region-, age-, and disease-specific values. No single published reference set spans all 16 basis functions at 3\,T, so the values of Table~\ref{tab:metabolites} were pooled across studies and are best read as a plausible, internally consistent choice rather than as consensus reference values. Limitations of this kind are common to synthetic \ac{mrs}, and documenting them explicitly is what allows a reader to judge whether the level of realism matches the intended use case \cite{LaMaster2026SyntheticReview, LaMaster2025MRSSim}.

\subsection{Future work}
\label{sec:future}
The limitations above map onto a concrete development path for the next release. Randomizing the metabolite concentrations and relaxation times per subject, and varying the noise level and the baseline parameters continuously, rather than in three discrete steps, would eliminate the concentration vectors that a \ac{dl} model can memorize. Releasing the macromolecular and baseline components with every partition, rather than with the testing ground truth alone, would allow nuisance signal removal and fitting to be scored separately on the training data as well. On the acquisition side, no outer-volume-suppression or lipid-nulling module is currently simulated. Automatic placement of suppression bands, together with Bloch simulation of the suppression pulses themselves, would alleviate the lipid-removal stage that most \textit{in vivo} pipelines rely on. A 7\,T version is a natural extension of this work. The subject anatomies were chosen so that 7\,T session data exist for the same individuals. Beyond these, subject motion, eddy currents, coil-sensitivity variation and imperfect excitation are the effects whose absence most limits transfer to \textit{in vivo} data, and the modular structure of the forward model admits any one of them without a redesign of the rest.

\section{Conclusions}
The 2024 \ac{mrsi} Data Processing and Quantification Challenge Synthetic Dataset presented here is a simulated 3\,T brain \ac{fid}--\ac{mrsi} resource with anatomical images, \(B_0\) maps, nuisance components, and ground-truth metabolite maps, together with a complete numerical specification of the forward model that generated it. The dataset is intended for reproducible development and evaluation of spectral processing, nuisance-signal removal, and metabolite-quantification methods. By documenting the simulation model, the exact contents of every component, the realized parameter ranges of all 32 datasets, and the points at which the implementation departs from its design, this work provides a foundation for a reusable community benchmark and a template for future synthetic \ac{mrsi} datasets.


\bmsubsection*{Author Contributions}

\textbf{John T. LaMaster:} Conceptualization, Methodology, Software, Validation, Formal analysis, Investigation, Resources, Data curation, Writing--original draft, Writing--review \& editing, Visualization.

\textbf{Julian P. Merkofer:} Conceptualization, Methodology, Software, Validation, Formal analysis, Investigation, Resources, Data curation, Writing--original draft, Writing--review \& editing, Visualization.

\textbf{Dennis M. J. van de Sande:} Conceptualization, Methodology, Software, Validation, Formal analysis, Investigation, Resources, Data curation, Writing--original draft, Writing--review \& editing, Visualization.

\textbf{Brian J. Soher:} Conceptualization, Methodology, Software, Validation, Formal analysis, Investigation, Resources, Data curation, Writing--review \& editing.

\textbf{Bernhard Strasser:} Conceptualization, Methodology, Software, Validation, Formal analysis, Investigation, Resources, Data curation, Writing--review \& editing.

\textbf{Chao Ma:} Conceptualization, Methodology, Software, Validation, Formal analysis, Investigation, Resources, Data curation, Writing--review \& editing, Supervision, Project administration.

\bmsubsection*{Financial Disclosure}

None reported.

\bmsubsection*{Ethics Statement}

Institutional review board statement: not applicable. Informed consent: not applicable.

\bmsubsection*{Data Availability Statement}

The data repository is available at Zenodo: \url{https://doi.org/10.5281/zenodo.21890221}.

\bmsubsection*{Acknowledgments}

We would like to thank Dr. Diana Waldmannstetter for her help with the image registration portion of our data generation pipeline, the Duke Department of Radiology for hosting the data the development and fitting challenge, and Drs. Ovidiu Andronesi and Wolfgang Bogner for their help in initiating the challenge that led to this dataset.

Data were provided in part by the Human Connectome Project, WU-Minn Consortium (Principal Investigators: David Van Essen and Kamil Ugurbil; 1U54MH091657) funded by the 16 NIH Institutes and Centers that support the NIH Blueprint for Neuroscience Research; and by the McDonnell Center for Systems Neuroscience at Washington University.

\bmsubsection*{Conflicts of Interest}

The authors declare no conflicts of interest.


\bibliography{references}

\clearpage
\input{SI}


\end{document}

%% file: SI.tex






\appendix

\section{The Basis Set}
\label{app:extra}

\begin{figure*}[p]
\centering
\input{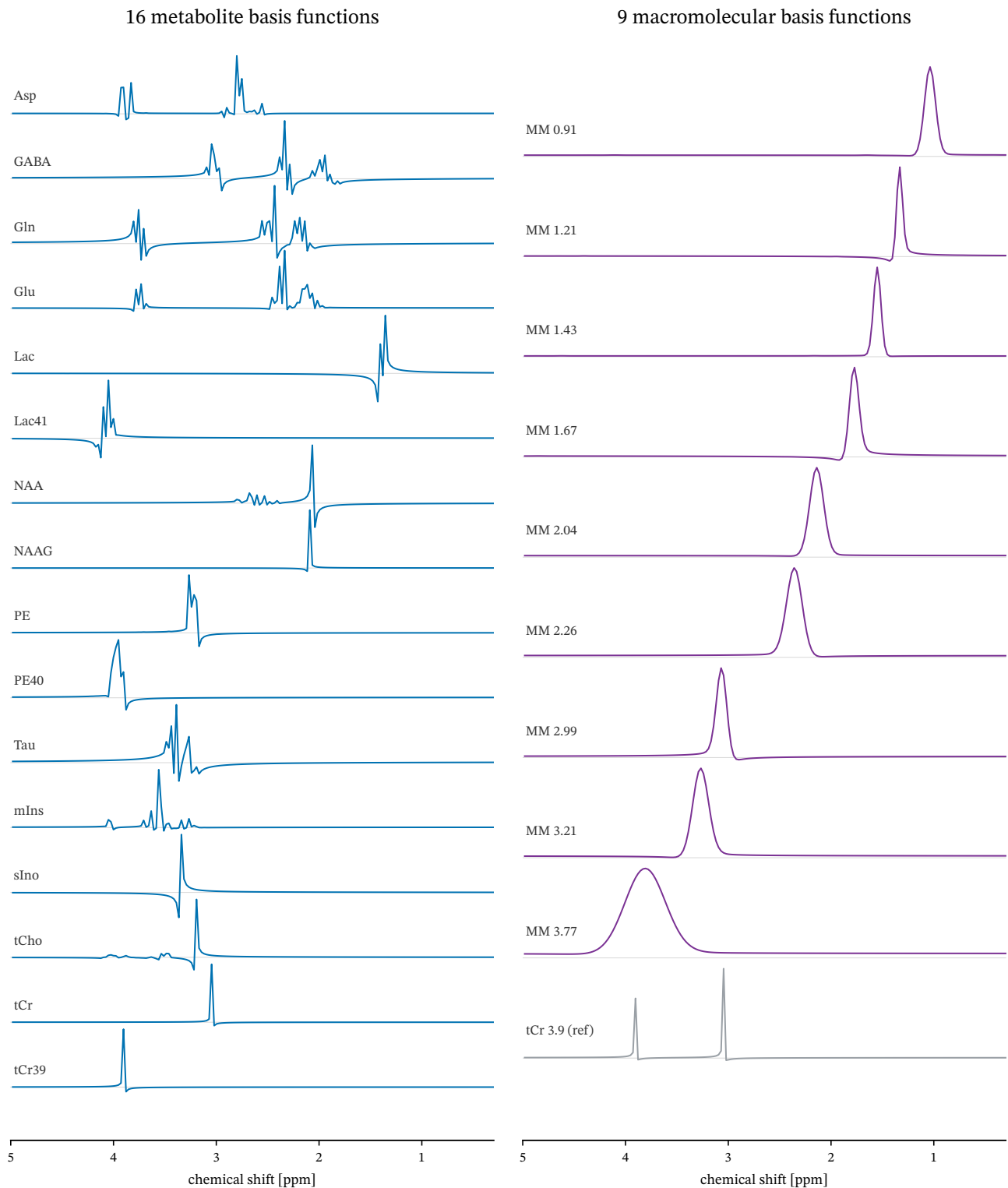}
\caption{The 16 metabolite (left) and 9 macromolecular (right) time-domain basis functions are displayed as spectra over $0.3$--$5$\,ppm without apodization. Each trace is normalized to its own tallest resonance and phased to that resonance. The basis functions were truncated above $4.3$\,ppm, so the region above that is empty. The small oscillations adjacent to the strongest peaks are the truncation response of the 386-point basis, which carries no intrinsic $T_2$. The tenth trace in the right-hand column (gray) is not a macromolecular component, rather it is the damped creatine reference used to normalize the \ac{mm} amplitudes. It is not injected into the signal. The metabolite basis functions labeled tCr39, PE40 and Lac41 hold the resonance groups of tCr, PE and Lac at or above $3.9$\,ppm. They were split off because they are differentially affected by water suppression.}\label{fig:basis}
\end{figure*}

Figure~\ref{fig:basis} shows all 26 time-domain basis functions distributed with the dataset, displayed as spectra. The resonance positions provide an independent check on the frequency axis: NAA appears at $2.01$\,ppm, total creatine at $3.03$ and $3.91$\,ppm, the lactate doublet at $1.33$\,ppm, and each macromolecular component at the chemical shift in its name.

\section{The MATLAB Convenience File}
\label{app:matfile}

Each subject is released with a single MATLAB v7.3 (HDF5) file holding every array of Table~\ref{tab:nifti} together with the basis functions and the sampling times, so that the whole dataset can be loaded in one call. Its contents are listed in Table~\ref{tab:contents}. It is redundant with the \ac{nifti} release, and is provided so that the whole of a subject's data, including the basis functions, can be read without assembling it from the individual volumes.

\begin{table*}[p]
\caption{Contents of the released per-subject \texttt{.mat} files. Training subjects (\texttt{Sub1}--\texttt{Sub24}) include the first block. The testing ground truth additionally includes \texttt{xtMM} and \texttt{xtBaseline}, which allow every term of Equation~\eqref{eq:overall_signal_model} to be isolated. Testing datasets for sub-challenge~2 do not contain \texttt{xtNuisance}.}\label{tab:contents}
\newcolumntype{L}{>{\raggedright\arraybackslash}X}
\begin{tabularx}{\textwidth}{llL}
\toprule
\textbf{Variable} & \textbf{Size} & \textbf{Content}\\
\midrule
\texttt{xtAll}      & $64\times64\times32\times384$ & Composite data: metabolites $+$ \ac{mm} $+$ nuisance $+$ baseline $+$ noise\\
\texttt{xtMeta}     & $64\times64\times32\times384$ & Noiseless metabolite signal $s_M$ \emph{only} (no \ac{mm})\\
\texttt{xtNuisance} & $64\times64\times32\times384$ & Noiseless truncated water $+$ lipid signal\\
\texttt{metaMap}    & $64\times64\times32\times16$  & Ground-truth metabolite amplitudes $a_n(\mathbf{r})$, Equation~\eqref{eq:metamap}\\
\texttt{VtMeta}     & $386\times16$                 & Metabolite time-domain basis functions $\varphi_n$\\
\texttt{VtMM}       & $386\times10$                 & 9 \ac{mm} basis functions $\varphi_{B,n}$, plus the tCr reference in row 10\\
\texttt{B0map}      & $64\times64\times32$          & Field offset $\Delta f$ in Hz, referenced to brain-mean water\\
\texttt{brainMask}  & $64\times64\times32$          & Binary brain mask\\
\texttt{Iref}       & $64\times64\times32$          & $T_1$-weighted image on the \ac{mrsi} grid\\
\texttt{IrefHr}     & $128\times128\times64$        & $T_1$-weighted image on the high-resolution grid\\
\texttt{t}          & $384\times1$                  & Sampling times, starting at \ac{te}\\
\texttt{hzpppm}, \texttt{ppmoff} & scalars          & $123.24$\,Hz/ppm and $4.70$\,ppm display offset\\
\midrule
\texttt{xtMM}       & $64\times64\times32\times384$ & Noiseless macromolecular signal $s_B$ (testing ground truth only)\\
\texttt{xtBaseline} & $64\times64\times32\times384$ & Noiseless baseline signal $s_{\mathrm{BL}}$ (testing ground truth only)\\
\bottomrule
\end{tabularx}
\end{table*}

\section{Implementation Details and Processing Chain}
\label{app:implementation}

This appendix records the implementation details so that the dataset can be regenerated. The pipeline is a sequence of five stages, executed once per subject.

\begin{enumerate}
    \item \textbf{Anatomical preparation.} HCP $T_1$-weighted (\texttt{*\_T1w\_MPR1}) and $B_0$ field map magnitude/phase volumes are unzipped and read. SPM8 ''New Segment'' is run against the \texttt{toolbox/Seg/TPM.nii} tissue-probability maps, yielding six probabilistic classes. The $B_0$ field map phase is interpolated onto the MPRAGE grid by linear \texttt{griddata}. The slice range is chosen as the last 183 slices with signal, giving a 128\,mm slab.
    \item \textbf{Water synthesis.} The tissue maps and $T_1$-weighted images are resampled to $128\times128\times64$. The $B_0$ field map is converted to Hz and the mean over the brain mask is removed. For every high-resolution voxel and every tissue class, $T_2$ is drawn from $\mathcal{N}(\mu_\tau, 5^2)$\,ms and $T_1$ from a uniform integer neighborhood of $\mu_\tau$, and both the unsuppressed and the \ac{wet}-suppressed water \ac{fid}s are accumulated using the tabulated $M_z^{\mathrm{WET}}(z, \Delta f, T_1)$. Both volumes are then Fourier-cropped to $64\times64\times32$.
    \item \textbf{Basis preparation.} The 16 metabolite and 9 macromolecular basis functions, together with the tCr reference, are read from their high-spectral-resolution \ac{nifti} representations. The metabolite bases are normalized by the peak of tCr39, and the \ac{mm} bases are normalized so that its 3.9\,ppm creatine reference matches the tCr39 basis damped with a 300\,ms time constant. Both are interpolated onto the acquisition raster.
    \item \textbf{Metabolite and \ac{mm} synthesis.} Tissue maps are resampled to $64\times64\times32$. For every voxel, the excitation profile of the \acs{slr} slab pulse is interpolated to the slice position. The steady-state weight of Equation~\eqref{eq:ernst} is evaluated separately for $T_1^M=1300$\,ms and $T_1^{MM}=250$\,ms, and Equations~\eqref{eq:metabolite_signal} and \eqref{eq:background_signal} are accumulated. The ground-truth map of Equation~\eqref{eq:metamap} is stored alongside the simulations. 
    \item \textbf{Lipid synthesis and assembly.} Lipid \ac{mrsi} data are registered into subject space (Section~\ref{sec:lipid}) using the registration performed by \texttt{antsRegistrationSyN}, which consists of rigid, affine, and \ac{syn} stages. The rigid and affine stages use mutual information with 32 histogram bins and $25\%$ regular sampling, whereas the \ac{syn} stage uses neighborhood cross-correlation with a radius of 4 and a gradient step of 0.1. The donor image is fixed and the subject image is moving during registration. The resulting transformation is inverted to map the donor image into the subject space and is applied to each time point in the \texttt{xt}-dimension using linear interpolation. The registered lipid data are Fourier-cropped to the target matrix, and added to the truncated water volume to form \texttt{xtNuisance}. The baseline is generated inside the brain mask with the parameters of Section~\ref{sec:baseline}, complex Gaussian noise is drawn, and all terms are summed into \texttt{xtAll}.
\end{enumerate}

The spatial preprocessing is fixed by the source geometry. The $256\times320\times320$ MPRAGE and its segmentation are cropped along the slice direction to the 183 source slices spanning the $128$\,mm excitation slab. They are then resampled by cubic-spline interpolation to the $128\times128\times64$ high-resolution grid on which the water and lipid components are synthesized. The same protocol then resamples original volumes to the $64\times64\times32$ target grid on which the metabolite, \ac{mm}, and baseline terms are built. The nominal voxel size of $2.8\times3.5\times4.0$\,mm$^3$ follows from this sequence of spatial transformations rather than from any file header.

Two implementation details are worth mentioning because they are easy to get wrong. The k-space cropping is performed with a centered rectangular window, indices $N/2+1-N_{\mathrm{lr}}/2$ through $N/2+N_{\mathrm{lr}}/2$ in each spatial dimension. The forward and inverse spatial FFTs from Equation~\eqref{eq:kspace_truncation} are normalized by $1/\sqrt{N_xN_yN_z}$. The tissue maps are permuted and flipped along all three axes when moving from the SPM output to the simulation space.

\section{Complete MRSsynMRS Reporting Checklist}
\label{app:checklist}
\setcounter{table}{0}

Table~\ref{tab:checklist} completes the proposed \ac{mrssynmrs} reporting table \cite{LaMaster2026SyntheticReview} for this dataset.

\clearpage
\onecolumn
\vspace*{\fill}
\begin{center}
\captionof{table}{\ac{mrssynmrs}-based reporting checklist for the 2024 \ac{mrsi} Challenge synthetic dataset.}\label{tab:checklist}
\begin{longtable}{@{}l p{0.635\textwidth}@{}}
\toprule
\textbf{Item} & \textbf{Value}\\
\midrule
\endfirsthead
\multicolumn{2}{@{}l}{\textit{Table \ref{tab:checklist}, continued}}\\
\toprule
\textbf{Item} & \textbf{Value}\\
\midrule
\endhead
\midrule
\multicolumn{2}{r@{}}{\textit{continued on the next page}}\\
\endfoot
\bottomrule
\endlastfoot
\multicolumn{2}{l}{\noindent \quad \textit{Purpose and scope}}\\
Intended use & Benchmarking of \ac{mrsi} nuisance removal and metabolite quantification; ML/DL training with the caveats of Section~\ref{sec:limitations}\\
Modality & 3D \ac{fid}--\ac{mrsi}, single voxel-wise coil combination assumed\\
\midrule
\multicolumn{2}{l}{\noindent \quad \textit{Acquisition-like parameters}}\\
Field strength, nucleus & 3\,T, $^1$H, $123.24$\,MHz\\
\ac{tr}, \ac{te} & $450$\,ms, $1.66$\,ms\\
Spectral width, points & $1204.8$\,Hz, 384\\
Spatial matrix, voxel size & $64\times64\times32$, $2.8\times3.5\times4.0$\,mm$^3$ (derived)\\
\midrule
\multicolumn{2}{l}{\noindent \quad \textit{Basis set}}\\
Simulator, version & Vespa-Simulation v1.1.1rc1 (PyGAMMA)\\
Sequence, \ac{te} used for basis & Ideal one-pulse ($90^\circ$), $\mathrm{TE}=0$\,ms\\
Number of metabolites & 16 (14 distinct; Lac and Lac41 have zero concentration)\\
Linewidth applied at simulation & None (no $T_2$ decay)\\
\midrule
\multicolumn{2}{l}{\noindent \quad \textit{Signal model}}\\
Amplitude model & Fixed \ac{gm}/\ac{wm} concentrations $\times$ tissue fraction $\times$ steady-state weight (Table~\ref{tab:metabolites})\\
Metabolite ranges & None: concentrations are constants, identical across all datasets\\
Relaxation & $T_1 = 1300$\,ms (all metabolites); per-metabolite $T_2$ (Table~\ref{tab:metabolites})\\
Lineshape & Voigt: per-metabolite Lorentzian $T_{2,n}$ plus a voxel-wise exponential term $\exp[-t/T_2'-t^2/T_G^2]$ (Equation~\eqref{eq:voigt})\\
Frequency shift & $\Delta f(\mathbf{r})$ from the subject $B_0$ map; no additional random shift\\
Line-shape parameters & $T_2'$ median $112$--$132$\,ms, $1/T_2'$ correlates with $|\nabla B_0|$ at $r=0.84$; $T_G = 114.0\pm2$\,ms\\
Phase & Determined by $\Delta f$ and \ac{te}; no random phase\\
\midrule
\multicolumn{2}{l}{\noindent \quad \textit{Nuisance components}}\\
\ac{mm} & 9 \textit{in vivo} \ac{hlsvd} components, amplitudes tied to tCr (Table~\ref{tab:mm}); measured per-component widths plus voxel-wise field broadening (Section~\ref{sec:mm})\\
Residual water & Bloch-simulated \ac{wet} (one $50$\,Hz pulse); tissue-specific $T_1$, $T_2$ with per-voxel jitter (Table~\ref{tab:water})\\
Lipid & Measured \textit{in vivo} at 3\,T, in-house \ac{fid}--\ac{epsi}, TR/TE $=300/4.7$\,ms, matrix $128\times128\times64$, field of view $240\times240\times192$\,mm$^3$; \ac{syn}-registered per subject, unsuppressed\\
Baseline & Bounded smoothed random walk, $1$--$6$\,ppm at 50 points/ppm, step $\sigma=0.05$, bounds $[-1,1]$, scale $0$--$0.2$; brain-masked, drawn per voxel; $12$--$17\%$ of the metabolite spectral peak\\
Noise & i.i.d.\ complex Gaussian, $\sigma \in \{0.6,1.0,1.4\}\times10^{-3}$ per channel\\
\midrule
\multicolumn{2}{l}{\noindent \quad \textit{Spatial information}}\\
Anatomical source & HCP 1200 release, $T_1$w MPRAGE $0.7$\,mm, gradient-echo field map ($\Delta\mathrm{TE}=2.46$\,ms)\\
Segmentation & SPM8 ``New Segment'', 6 classes\\
$B_0$ map & Subject-specific, demeaned over the brain; range in Table~\ref{tab:partitions}\\
$B_1$ map & Not modeled (uniform transmit/receive assumed)\\
k-space model & Rectangular crop, $128^2\times64 \rightarrow 64^2\times32$, applied to water and lipid only\\
\midrule
\multicolumn{2}{l}{\noindent \quad \textit{Dataset and access}}\\
Partitions & 24 training; 5 testing (sub-challenge 1); 3 testing (sub-challenge 2)\\
Ground truth provided & Metabolite maps and signals; nuisance signals; \ac{mm} and baseline for the testing ground truth only\\
Format & \ac{nifti}, \ac{nifti-mrs}, MATLAB \texttt{.mat} (v7.3)
\end{longtable}
\end{center}
\vspace*{\fill}
\twocolumn

